\documentclass[12pt]{article}

\usepackage{latexsym}
\usepackage{cite}
\usepackage{bm}
\usepackage{relsize}
\usepackage{geometry}
 \usepackage{amsmath, amssymb, amscd, xypic, graphicx}
\usepackage{slashed,color}
\usepackage{epstopdf}
\usepackage[colorlinks]{hyperref}
\usepackage[displaymath, tightpage]{preview}
\input xy
\xyoption{all}
\def\beq{\begin{equation}}
\def\eeq{\end{equation}}
\def\beqn{\begin{eqnarray}}
\def\eeqn{\end{eqnarray}}

\newcommand{\nzt}{${\mathcal N}\!\!=\!(0,2)\,$}

\newcommand{\cf}{${\mathcal F}$}
\newcommand{\cfe}{{\mathcal F}}

\newcommand{\sma}{\left(\begin{smallmatrix}}
\newcommand{\smaa}{\end{smallmatrix}\right)}

\newcommand{\gsim}{\lower.7ex\hbox{$
\;\stackrel{\textstyle>}{\sim}\;$}}
\newcommand{\lsim}{\lower.7ex\hbox{$
\;\stackrel{\textstyle<}{\sim}\;$}}

\newcommand*\xbar[1]{%
 \kern0.5ex%
  \hbox{%
   \kern0.2ex%
      \vbox{%
      \hrule height 0.5pt 
      \kern0.5ex
      \hbox{%
        \kern-0.1em
        \ensuremath{#1}%
        \kern-0.1em
      }%
    }%
  }%
}

\renewcommand{\theequation}{\thesection.\arabic{equation}}

\begin{document}

\begin{titlepage}

\begin{flushright}
FTPI-MINN-14/5, UMN-TH-3326/14,  NSF-KITP-14-025\\[2mm]
\end{flushright}

\vspace{1mm}

\begin{center}
{  \Large \bf  {\boldmath${\mathcal N}\!=\! (0,2)$} Deformation of {\boldmath$(2,2)$}
Sigma Models: \\[1mm]
Geometric Structure, Holomorphic Anomaly \\[3mm]
and Exact
 {\boldmath$\beta$} Functions }
\end{center}

\vspace{0.5cm}

\begin{center}
{\large
Jin Chen,$^{a}$ Xiaoyi Cui,$^{b}$ Mikhail Shifman,$^{a,c}$ \\[2mm] and Arkady Vainshtein$^{\,a,c,d}$}
\end {center}

\vspace{1mm}

\begin{center}

$^{a}${\it  Department of Physics, University of Minnesota,
Minneapolis, MN 55455, USA}\\[1mm]
$^{b}${\it Max Planck Institute for Mathematics, Bonn, 53111, Germany}\\[1mm]
$^c${\it  William I. Fine Theoretical Physics Institute,
University of Minnesota,
Minneapolis, MN 55455, USA}\\[1mm]
$^{d}${\it Kavli Institute for Theoretical Physics, University of California,Santa Barbara, CA 93106, USA}

\end{center}

\vspace{0.6cm}

\begin{center}
{\large\bf Abstract}
\end{center}

We study  $\mathcal N\!=\!(0,2)$ deformed (2,2) two-dimensional sigma models. Such hete\-rotic models were
discovered previously on the world sheet of non-Abelian strings supported by certain four-dimensional $\mathcal N\!=\!1$ theories.
We study geometric aspects and holomorphic properties of these models, and derive a number
of exact expressions for the $\beta$ functions in terms of the anomalous dimensions analogous to the NSVZ $\beta$ function in four-dimensional gauge theories. Instanton calculus provides a straightforward  method for the derivation verified at two loops by explicit calculations. 
Obtained two-loop anomalous dimensions give an explicit expression for one of the $\beta$ functions 
valid to three loops.
The fixed point in the ratio of the couplings found previously at one loop is not shifted  at two loops.
We also consider the $\mathcal N\!=\!(0,2)$ supercurrent supermultiplet (the so-called hypercurrent) and its anomalies, as well as the ``Konishi anomaly."
This gives us another method for finding exact $\beta$ functions.
We prove that despite the chiral nature of the models under consideration quantum loops preserve 
isometries of the target space. 

\end{titlepage}

\newpage

\tableofcontents

\newpage

\section{Introduction}
\label{intro}
\setcounter{equation}{0}

Heterotically deformed $\mathcal{N}\!=(0,2)$ sigma models to be considered below
emerged as low-energy world sheet theories on non-Abelian strings supported in some $\mathcal{N}\!=1$
four-dimensional Yang-Mills theories \cite{Edalati:2007vk, Shifman:2008wv} (for a recent review see  \cite{Shifman:2014jba}).
In this case in particular the target space was CP($N\!-\!1$) but the heterotic modification could be considered
for a wide class of the K\"ahler manifolds. The heterotic models, although
remaining largely unexplored, appear to be important in various problems. Some previous results can be found in
\cite{Cui:2011rz,Cui:2010si,Cui:2011uw,Shifman:2008kj,B1}, for a general discussion of (0,2) models see \cite{Witten:2005px,bai2,bai3,Adams:2003zy,Melnikov:2012hk}. A renewed interest is also due to the recent publications \cite{Jia,Gadde:2013lxa,Gadde:2014ppa}.
Here we report further results in
the study of geometric structure, holomorphic anomaly and exact $\beta$ functions
in these models.

Heterotic $\mathcal{N}\!=(0,2)$ models have a rich mathematical structure.
 In perturbation theory two-dimensional
  $\mathcal{N}\!=\!(0,2)$
models were shown to share some features  with $\mathcal{N}\!=\!1$ super-Yang-Mills theories in four dimensions
(e.g. \cite{Cui:2011uw}). They are asymptotically free in the ultraviolet (UV), have different phases
in the infrared (IR), and admit large-$N$ solution \cite{Shifman:2008kj,B1}. These
facts can be interpreted within  4D--2D correspondence  (e.g. \cite{Edalati:2007vk, Shifman:2008wv}) and
 the Dijkraaf-Vafa-type deformation. The same correspondence was noted  in  more general 4D--2D coupled systems,
the study of which requires a tho\-rough  knowledge of the two-dimensional side. A number of insights were obtained from string theory, see \cite{Adams:2003zy, Melnikov:2012hk}. However, considerations of  (0,2) models in quantum field theory are scarce. In particular, beyond chiral operators, nothing was explored until quite recently.

Two-dimensional sigma models present a natural playground for geometric explorations.\ They encode the geometry of the target space, that of the world sheet, and the geometry of various moduli spaces. Essentially everything is known for the undeformed $\mathcal{N}\!=\!(2,2)$ models. With $\mathcal{N}\!=\!(0,2)$ supersymmetry, one can test the robustness of
the target space geometry -- whether or not quantization provide us with some kind of geometrical deformation. In particular, the models we will consider have both, isometries and a global symmetry realized in a nontrivial way. The interplay between geometry and quantum effects could be enlightening on both sides.

Finally, implications of current algebra in $\mathcal{N}\!=\!(0,2)$ theories were discussed more than once. While the general structure is known \cite{Dumitrescu:2010ca, Dumitrescu:2011iu}, explicit examples of how these
current-algebraic relations are implemented in particular models and what they imply for quantization were not worked out. We emphasize that the current algebra calculation and the renormalization group (RG) flow of the theory are intertwined \cite{shiva}, and, hence, it is possible to formulate the current algebra as a way to uncover renormalization of a given theory. Moreover, the
supercurrent supermultiplet (to be referred to as {\em hypercurrent}) starts from the U(1)$_R$ current; therefore the overall anomaly is determined by the index theorem for the appropriate Dirac operator.

To explain the nature of heterotic modifications let us start with reminding geometry of unmodified $\mathcal{N}\!=\!(2,2)$ sigma models.
It was pointed out by Zumino \cite{Zumino:1979et} that the target space of these models should have the K\"ahler geometry.
Moreover, to be characterized  just by one coupling $g$, it should be a symmetric space which can be described as a homogeneous space $G/H$
for a Lie group $G$ and the stabilizer $H$. For the projective CP($N\!-\!1$) space
$G={\rm SU}(N)$ and $H={\rm S}\left({\rm U}(N\!-\!1) \times  {\rm U}(1)\right)$.
It is a particular case of Grassmannian spaces $${\rm SU}(n+m)/{\rm SU}(n)\times {\rm SU}(m)\times {\rm U}(1)$$ (see, e.g., \cite{Morozov:1984ad} for a full list of the symmetric K\"ahler spaces).

In these homogeneous spaces the Ricci tensor $R_{i\bar j}$ is proportional to
the metric $G_{i\bar j}$\,,
\beq
R_{i\bar j}=b \, \frac{g^{2}}{2}\,G_{i\bar j}\,.
\eeq
This feature is a definition of the  Einstein spaces. The constant $b$
is equal to the dual Coxeter number $T_{G}$ for the group $G$.,
\beq
b_{\,G/H}=T_{G}\,.
\eeq
Correspondingly, for the CP($N\!-\!1$) space
\beq
b_{\,{\rm CP}(N-1)}=T_{{\rm SU}(N)}=N\,.
\eeq
The same constant $b=T_{G}$ defines the  $\beta$ function
\beq
\beta_{{\cal N}\!=\!(2,2)}(g)=\mu\,\frac{dg^{2}}{d\mu}=-T_{G}\,\frac{g^{4}}{4\pi}\,,
\eeq
which is exhausted by one loop \cite{Novikov:1983mt, Morozov:1984ad} in the (2,2) theories.

To diminish the number of supercharges from 4 in $\mathcal{N}\!=\!(2,2)$ to 2 in $\mathcal{N}\!=\!(0,2)$
one needs to break partially a partnership between bosonic and fermionic fields. In the (2,2) case 
each bosonic
field $\phi^{i}$ has two fermion partners, right- and left-movers, $\psi_{R}^{i}$ and $\psi_{L}^{i}$\,.
A simple way to diminish supersymmetry to (0,2) is just to discard all $\psi_{R}^{i}$\,. Such chiral models, which can be called
{\em minimal}, generically suffer from internal diffeomorphism anomaly \cite{Moore:1984ws,Bagger:1985pw}. 
The existence of this anomaly is due to the fact that the first Pontryagin class $p_1$ of the manifold does not vanish.
The  only exception from the entire CP($N\!-\!1$) series is CP(1) which is free from this anomaly having
vanishing $p_1$. On the other hand, CP(1) has nonvanishing first Chern class $c_1$ which leads to a problem
in its gauged formulation. Still, the theory can be made consistent by deleting one point in its target 
manifold \cite{NikNek}.\footnote{\,We are indebted to Nikiita Nekrasov for explaining these subtle points.} 
See also the discussion in \cite{hull}. In this paper we do not consider minimal (0,2) models.

In the approach to $\mathcal{N}\!=\!(0,2)$ gauge theories in 2D developed in
In Refs.\cite{Gadde:2013lxa,Gadde:2014ppa} $\mathcal{N}\!=\!(0,2)$ the matter content was chosen
in a way that leads to cancellation of anomalies.

Our study is focused on a different heterotic modification. Namely, instead of deleting
right-moving fermions $\psi_{R}^{i}$, one extra right-mover $\zeta_{R}$ is added to the content
of the (2,2) theory. A new coupling which mixes $\zeta_{R}$ and $\psi_{R}^{i}$ in the background
of the bosonic field leads to breaking of the (2,2) supersymmetry to (0,2). In contrast to the minimal (0,2)
models this heterotic coupling can be switched on perturbatively. Together with the singlet nature 
of the extra field $\zeta_{R}$ it is sufficient to show the absence of the internal anomaly problem.

Thus, our task in the present paper is to analyze the heterotically deformed (nonminimal)
$\mathcal{N}\!=\!(0,2)$ models.
We present a more complete geometric formulation of the class of nonminimal models
which will be studied in this paper. Holomorphic properties of such models are
revealed. They have two coupling constants, the original $g$ and the heterotic $h$\,. 
Correspondingly, there are two $\beta$ functions. As in four-dimensional Yang-Mills \cite{shiva}, 
we have to differentiate between
the holomorphic coupling constants which are renormalized at most at one loop,
and their nonholomorphic counterparts. The latter appear in conventional perturbation theory and 
are sometimes referred to as canonic.

We calculate both $\beta$ functions in more than one way. In particular, we derive exact relations between the $\beta$ functions
and the anomalous dimensions $\gamma$, analogous to the NSVZ relations in four-dimensional 
${\cal N}\!=\!1$ gauge theories \cite{Novikov} using the instanton calculus. For instance, for $\beta_g$ it will be shown that to all orders in perturbation theory
\beq
\beta_{g}\!=\!\mu\,\frac{dg^{2}}{d\mu}\!=\!-\frac{g^{2}}{4\pi}\,\frac{T_{G}\,g^{2}\left(1+\gamma_{\psi_R}/2
\right) - {h}^{2}\left(\gamma_{\psi_{R}}+\gamma_{\zeta}\right)}{1-({h}^{2}/4\pi)}\,,
\label{11}
\eeq
where $\gamma_{\psi_{R}},\,\gamma_{\zeta}$ are the anomalous dimensions of the $\psi_{R},\,\zeta_{R}$ fields.

We compute the anomalous dimensions up to two loops implying prediction for three loops in $\beta_g$\,. Our two-loop results for anomalous dimensions also confirm the fact that there exists a fixed point for the ratio $\rho\equiv h^2/g^2 $. The critical value
$\rho_{c}$ depends  on a manifold geometry and equals to 1/2 for CP($N\!-\!1$)\cite{Cui:2010si}. At this point three-loop $\beta_g$ reduces to
\beq
\beta^{(3)}_{g}\Big|_{\rho=\rho_c}\!=\!-T_{G}\,\frac{g^{4}}{4\pi}\,\frac{1}{\big(1-(h^{2}/4\pi)\big)}\,,
\label{11a}
\eeq
i.e., to the one-loop expression up to the factor $1/(1-(h^{2}/4\pi))$.

Then we prove that despite the chiral nature of the model all (2,2) isometries of the target space
are preserved by the heterotic (0,2) modification. A feature which differentiate the model from other (0,2)
theories and allows for a simple proof is that the extra chiral fermion is a singlet of the the group $G$ in 
the symmetric $G/H$ space.

Finally, we consider the $\mathcal N\!=\!(0,2)$ supercurrent supermultiplet (hypercurrent) and its anomalies,
as well as the ``Konishi anomaly" \cite{KK}.
This gives us another method for finding the exact expression for $\beta_{g}$ via anomalous dimensions.

\section{Heterotic \boldmath{${\mathcal N}=(0,2)$} models in 2D}
\label{loopcalc}
\setcounter{equation}{0}

\subsection{Formulation of the model}

We start to describe the model to be studied in this paper by introducing two types of chiral $\mathcal{N}\!=\!(0,2)$ superfields $A$ and $B$.
The first, bosonic superfield $A$ describes a chiral supemultiplet which on mass shell consists of
a complex bosonic field and a left-moving Weyl fermion,
\beq
A(x_{R}+2i\theta^\dagger\theta,x_{L},\theta)=\phi(x_{R}+2i\theta^\dagger\theta,\,x_{L})+\sqrt{2}\,\theta\,\psi_L(x_{R}+2i\theta^\dagger\theta,\, x_{L})\,.
\label{15}
\eeq
Here $x_{R,L}$ are light-cone coordinates $x_{R,L}=t\pm x$ and $\theta$ is the one-component Grassmann variable corresponding to $\theta_{R}$ (see Appendix A for our notation).
The second, fermonic superfield $B$ refers to the Fermi supermultiplet which on mass shell contains only the right-moving fermion ($F$ is an auxiliary
field),
\beq
B(x_{R}+2i\theta^\dagger\theta,x_{L},\theta)=\psi_R(x_{R}+2i\theta^\dagger\theta,\, x_{L})+\sqrt{2}\,\theta F(x_{R}+2i\theta^\dagger\theta,\, x_{L})\,.
\label{15-1}
\eeq
Note that in the nonlinear formulation here, the fermionic multiplets are taken to be chiral in a strict sense. In the $\mathcal{N}\!=\!(0,2)$ gauged formulation this condition can usually be relaxed. Note also that the  $\mathcal{N\!}=\!(2,2)$ chiral field $\Phi(x_{R}\!+\!2i\theta_{R}^{\dagger}\theta_{R},x_{L}\!-\!2i\theta_{L}^{\dagger}\theta_{L},\theta_{R},\theta_{L})$ decomposes
in the   $\mathcal{N}\!=\!(0,2)$ superfields $A$ and $B$ as
\beq
\begin{split}
\Phi(x_{R}+2i\theta_{R}^{\dagger}\theta_{R},x_{L}-2i\theta_{L}^{\dagger}\theta_{L},\theta_{R},\theta_{L})\qquad \qquad\qquad\qquad ~~~\\[2mm]
=A(x_{R}+2i\theta_{R}^\dagger\theta_{R},x_{L}-2i\theta_{L}^{\dagger}\theta_{L},\theta_{R})+\sqrt{2}\,\theta_{L} B(x_{R}+2i\theta_{R}^\dagger\theta_{R},x_{L},\theta_{R}).
\end{split}
\label{decomp}
\eeq
The $\mathcal{N}=(0,2)$ supersymmetry transformations are as follows:
\beqn
&& \delta x_{R}=-2i \theta^{\dagger}\epsilon\,, \quad  \delta x_{L}=0\,, \quad \delta\theta=\epsilon\,,\quad
\delta \theta^{\dagger}=\epsilon^{\dagger}\,,\nonumber\\[3mm]
&&\delta\phi=\sqrt{2}\,\epsilon \,\psi_L\,,\qquad \delta\psi_L=-\sqrt{2}\,i\epsilon^\dagger \partial_L\phi\,,\\[3mm]
&&\delta\psi_R=\sqrt{2}\,\epsilon \,F\,,\qquad \delta F=-\sqrt{2}\,i\epsilon^\dagger\partial_L\psi_R\,,
\nonumber
\label{n02transformation}
\eeqn
where $\partial_{L}=\partial_{t}+\partial_{x}=2\,\partial_{x_{R}}$.

The undeformed $\mathcal{N}\!=\!(2,2)$
model in terms of the $\mathcal{N}\!=\!(0,2)$
superfields (\ref{15}) and (\ref{15-1})  contains equal number of bosonic $A^{i}$ and fermionic
$B^{i}$ superfields. In the particular case of CP($N\!-\!1$) we have  $i=1,2,...,N-1$.

The heterotic deformation to be considered below is induced
by adding a singlet fermionic superfield $\mathcal{B}$,
\beq
 \mathcal{B}=\zeta_R(x_{R}+2i\theta^\dagger\theta,\,x_{L}) +\sqrt{2}\,\theta\mathcal{F}(x_{R}+2i\theta^\dagger\theta,\,x_{L})\,.
 \eeq
The Lagrangian
can be written as
\beqn
\mathcal{L}\! &=& \!  \frac{1}{4} \!\int\! d^2 \theta\, \Big[
K_i(A, A^\dagger)\left( i{\partial_{R}} A^i - 2\kappa\, \mathcal{B}B^{i}\right)+{\rm H.c.}\Big] \nonumber\\[1mm]
&+&\! \frac{1}{2} \!\int\! d^2 \theta\, \Big[Z\, G_{i\bar j}(A, A^\dagger)\, B^{\dagger\bar j} B^i +{\mathcal Z} \,\mathcal{B}^\dagger \mathcal{B}\Big]\,.
\label{sfLagN}
\eeqn
Here $\kappa$ is the new coupling constant defining the heterotic modification, $K$ is the K\"ahler potential viewed as a function of the bosonic superfields. By definition
\beq
K_i(A, A^\dagger) \equiv \partial_{A^i} \, K(A, A^\dagger)\,.
\label{5a}
\eeq
Moreover, $G_{i\bar j}$ is the metric on the target space,
\beq
G_{i\bar j} = K_{i\bar j}(A, A^\dagger) \equiv \partial_{A^i} \partial_{A^{\dagger \bar j}}\, K(A, A^\dagger)\,.
\eeq
Two $Z$ factors (for the fields $B^{i}$ and $\mathcal B$) are introduced in (\ref{sfLagN}),
in anticipation of their renormalization group (RG) evolution. In the CP$(N\!-\!1)$ model
$$
K (A, A^\dagger)=\frac{2}{g^{2}}\,\log \Big(1+\sum_{i}^{N-1}A^{\dagger\, i}\,A^{i}\Big),
$$
see Eq.\,(\ref{212}) for the other metric objects.

One can check  that the above Lagrangian is target-space invariant.
However, the target-space invariance is implicit in Eq.\,(\ref{sfLagN}) because $K_i(A, A^\dagger)$ is not explicitly
 target-space invariant. It becomes explicit upon passing to the integration over the Grassmann half-space in the first line of
 Eq.\,(\ref{sfLagN}),
\beqn
\mathcal{L}
\!&=&\!  -\frac{1}{4} \int \!  d \theta\,  G_{i\bar j}(A, A^\dagger) (\xbar{\!D}A^{\dagger \bar j})\left(i\partial_{R} A^i -2\kappa \,\mathcal{B} B^i\right)+{\rm H.c.}
\nonumber\\[3mm]
&~&+
\frac{1}{2}\!\int \!d^2 \theta\left[
Z\,G_{i\bar j}(A, A^\dagger)\, B^{\dagger\bar j} B^i  +  {\mathcal Z} \,\mathcal{B}^\dagger \mathcal{B}\right].
\label{sfLa}
\eeqn
The F-term structure $G_{i\bar j}(\xbar{\!D}A^{\dagger \bar j})i\partial_{R} A^i$ in the first line is an analog of
that for
the gauge term $W^{\alpha}W_{\alpha}$ in  4D gauge theories, while another F-structure, $G_{i\bar j}(\xbar{\!D}A^{\dagger \bar j})\,\mathcal{B} B^i$,
is an analog of  superpotential in 4D. The chiral nature of these terms plays a crucial role in their renormalization.
Of course, the second line can also be written as an integral over the Grassmann half-space, so that
the Lagrangian takes the form
\beq
\begin{split}
\mathcal{L} ={\rm Re}\int \!  d \theta\, \cfe=\frac 1 2 \int \!  d \theta\, \cfe+{\rm H.c.}\,, \hskip 6.0cm\\[2mm]
 \cfe=-\frac 1 2\, G_{i\bar j}(\xbar{\!D}A^{\dagger \bar j})\left(i\partial_{R} A^i -2\kappa \,\mathcal{B} B^i\right)-\frac 1 2\, \xbar{\!D}\left[Z\,G_{i\bar j}\, B^{\dagger\bar j} B^i  +  {\mathcal Z} \,\mathcal{B}^\dagger \mathcal{B}\right]\,.
\end{split}
\label{LF}
\eeq
In the chiral superfield integrand \cf ~the second line of (\ref{sfLa}) produces the derivative term with $\xbar{\!D}$ which
is not protected under renormalization.

The target space invariance is also  transparent if one rewrites the Lagrangian in components,
\beqn
&&\hspace{-10mm}\mathcal{L} =
G_{i\bar j}\left[\partial_{R}\phi^{\dagger \bar j} \partial_{L}\phi^{i}+\psi_{L}^{\dagger \bar j}\,i\nabla_{\!R}\,\psi_{L}^{i}
+ Z\,\psi_{R}^{\dagger \bar j}\,i\nabla_{\!L}\psi_{R}^{i}\right]
 + Z\,R_{i{\bar j} k {\bar l}}\,\psi_{L}^{\dagger \bar j}\psi_{L}^{i} \,\psi_{R}^{\dagger \bar l}\psi_{R}^{k} \nonumber\\[2mm]
&&\hspace{-6mm}  +{\mathcal Z}\,\zeta_R^\dagger \, i\partial_L \, \zeta_R  +\!
\left[\kappa\, \zeta_R  \,G_{i\bar j}\big( i\,\partial_{L}\phi^{\dagger \bar j}\big)\psi_R^{i}
+{\rm H.c.}\right] \! +\frac{|\kappa |^2}{Z} \zeta_R^\dagger\, \zeta_R
\big(G_{i\bar j}\,  \psi_L^{\dagger \bar j}\psi_L^{i}\big)\label{components}\\[2mm]
&&\hspace{-6mm}-\,\frac{|\kappa |^2}{{\mathcal Z}}
 \big(G_{i\bar j}\psi_{L}^{\dagger \bar j}\psi_{R}^{i}\big) \big(G_{k\bar l}\psi_{R}^{\dagger \bar l}\psi_{L}^{k}\big)\,.\nonumber
\eeqn
Here $\nabla_{\!L,R}$ are covariant derivatives, $\nabla_{\!L,R}\,\psi_{R,L}^{i}=\partial_{L,R}\,\psi_{R,L}^{i}+\Gamma^{i}_{kl}\,\partial_{L,R}\,\phi^{k}\,\psi_{R,L}^{l}$\,. The first line in this equation refers to the undeformed
$\mathcal{N}\!=\!(2,2)$ theory, the subsequent terms bring in the (0,2) deformation.

Actually all the above equations are applicable to the heterotic deformation of any  K\"ahler manifold.
In the particular case of CP$(N\!-\!1)$ the explicit expression for
the Fubini-Study metric and related objects are of the form,
\beqn
&& K=\frac{2}{g^{2}}\,\log \chi\,,\qquad \qquad \qquad \qquad ~~~~~~\chi=1+\sum_{m}^{N-1}\phi^{\dagger\, m}\phi^{m}\,,\\
&& G_{i\bar j}=\frac{2}{g^{2}}\Bigg(\frac{\delta_{i\bar j}}{\chi}-\frac{\phi^{\dagger\,i}\phi^{\bar j}}{\chi^{2}}\Bigg)\,,\qquad
\qquad
~G^{i\bar j}=\frac{g^{2}}{2}\,\chi\,\Big(\delta^{i\bar j}+\phi^{i}\phi^{\dagger \,\bar j}\Big)\,, \nonumber\\[1mm]
&& \Gamma^{i}_{kl}=-\frac{\delta^{i}_{k}\,\phi^{\dagger\,l}+\delta^{i}_{l}\,\phi^{\dagger\,k}}{\chi}\,,
\qquad \qquad\quad ~\Gamma^{\bar i}_{\bar k \bar l}=-\frac{\delta^{\bar i}_{\bar k}\,\phi^{ \bar l}+\delta^{\bar i}_{\bar l}\,\phi^{ \bar l}}{\chi}\,,\nonumber \\[1mm]
&& R_{i{\bar j}k{\bar l}}=-\frac{g^{2}}{2}\Big(G_{i\bar j}G_{k\bar l}+G_{k\bar j}G_{i\bar l}\Big)\,,\quad
 ~R_{i\bar j}= - G^{k \bar j}R_{i{\bar j}k{\bar l}}=\frac{g^{2}N}{2}\, G_{i\bar j}\,.\nonumber
 \label{212}
\eeqn

\subsection{Geometry of heterotic deformation}
\label{geometry}

What is the geometrical meaning of the heterotic deformation?
For the K\"ahler manifold $M$ of the complex dimension $d$ (for ${\rm CP}(N\!-\!1)$ the dimension $d=N\!-\!1$) we have
$d$ right-moving fermions
$\psi^{i}_{R}\,,\,i=1,\ldots,d$, plus $\zeta_{R}$. They can be viewed as  defined on the tangent bundle $T(M\times C)$.
Let us denote $\zeta_{R}=\psi_{R}^{\,d+1}$. Similarly,
for superfields ${\cal B}=B^{\,d+1}$. Then, the Lagrangian for the right-moving fermions
can be written as
\beq
{\cal L}_{B}=\frac{1}{2}\!\int\! d^2 \theta \left\{G^{(B)}_{i\bar j}B^{\dagger\,\bar j}B^{i}
+\Big[T_{ik}B^{i}B^{k}+{\rm H.c.}\Big]\right\}\,, \quad (i,k,\bar j=1,\ldots, d+1).
\eeq
Here the metric $G^{(B)}_{i\bar j}$ and antisymmetric potential $T_{ik}$ are functions
of the bosonic superfields $A^{i},\,A^{\dagger \bar j}$ with $i,\bar j=1,\ldots, d$.
Comparing with the previous definitions we see that nonvanishing components of $G^{(B)}_{i\bar j}$
and $T_{ik}$ are
\beqn
&&G^{(B)}_{i\bar j}=\left\{\begin{tabular}{l}$Z G_{i\bar j}\,,\quad i,\bar j =1,\ldots, d,$ \\[2mm]
${\cal Z}\,, \quad\quad ~i=d+1, ~~\bar j=d+1\,, $\end{tabular}\right.
\label{GB}\\[3mm]
&& T_{(d+1)\,i}=-T_{i\,(d+1)}=-\frac{\kappa}{2}\,K_{i}\,,\quad i=1,\ldots, d\,.
\eeqn
The potential $T_{ik}$ is not uniquely defined but its curvature
\beq
{\cal H}_{ik\bar j}\!=T_{ik,\bar j}=\frac{\partial T_{ik}}{\partial A^{\dagger \,\bar j}}
\label{curv}
\eeq
is a good object. This curvature defines the chiral form for the heterotic modification,
\beq
{\cal L}_{B}=\frac{1}{2} \!\int\! d^2 \theta \,G^{(B)}_{i\bar j}B^{\dagger\,\bar j}B^{i}
-\frac{1}{2}\!\int\! d\theta \Big[{\cal H}_{ik\bar j} (\xbar{\!D}A^{\dagger \bar j})B^{i}B^{k}+{\rm H.c.}\Big]\,,
\label{Lb}
\eeq
In the model at hand the nonvanishing components of ${\cal H}_{ik\bar j}$
\beq
{\cal H}_{(d+1)\,i\bar j} =-{\cal H}_{i\,(d+1)\,\bar j}=-\frac{\kappa}{2} \,G_{i\bar j} \qquad (i,\bar j=1,\ldots, d \,)
\label{Hcomp}
\eeq
are expressed via the metric tensor $ G_{i\bar j}$. It looks even simpler for
${\cal H}_{ik}^{j}={\cal H}_{ik\bar j}G^{j\bar j}$\,:
\beq
{\cal H}_{(d+1)\, i}^{j} =-{\cal H}_{i\,(d+1)}^{j}=-\frac{\kappa}{2} \,\delta_{i}^{j} \qquad (i,j=1,\ldots, d )\,.
\eeq
The chiral field ${\cal F}$ in Eq.\,(\ref{LF}) which defines the total Lagrangian, ${\cal L}={\rm Re} \int d \theta {\cal F}$, can be rewritten in the following generic form:
\beq
\cfe=-\frac 1 2 \left[ i\,G_{i\bar j}(\xbar{\!D}A^{\dagger \bar j})\partial_{R} A^i +{\cal H}_{ik\bar j} (\xbar{\!D}A^{\dagger \bar j})B^{i}B^{k}+ \xbar{\!D}\big(G_{i\bar j}^{(B)}B^{\dagger\bar j} B^i \big)\right]\,.
\label{LF1}
\eeq

In differential geometry the heterotic construction presented above can be described as an action of the
$C^{*}$ one-dimensional algebra
on the odd tangent vector bundle.\footnote{We are indebted to Alexander Voronov for explanations.}
Note that there is a diagonal U(1) which rotates $B^{i}$ and ${\cal B}$ in the opposite directions.

The consideration above is applicable to modifying any K\"ahler manifold, we are not limited to CP($N\!-\!1$).
What is less clear, whether or not it is possible to add more right-moving fermions so that the extra fermionic bundle
is not just $TC$. To this end a three-form ${\cal H}$ consistent with the target-space  invariance should exist.
We are not aware of such generalizations.

\subsection{Holomorphy and its breaking}
\label{sec:holo}
The deformed (0,2) theory contains four bare parameters:
\beq
\frac{1}{g^{2}}\,,~\frac{\kappa}{g^{2}}\,,~Z\,,~{\mathcal Z}\,.
\eeq
The first two, $1/g^{2}$ and $\kappa/g^{2}$, enter as coefficients of the $F$ terms in Eq.\,(\ref{sfLa}) and can be taken to be complex, while parameters $Z$ and ${\mathcal Z}$ should be real. The imaginary  part of $1/g^{2}$ defines the vacuum $\theta$ angle,
${\rm Im}(1/g^{2})=\theta/4\pi$\,, while the phase of $(\kappa/g^{2})$ produces an addition to this $\theta$ angle. These angles do not contribute to physical effects due to the presence of massless fermionic fields whose phase can be redefined.

Nonrenormalization of superpotential (i.e. the $\kappa$ term in (\ref{sfLa})) implies the absence of loop corrections to the holomorphic
coupling $\kappa/g^{2}$, and, in particular, the absence of its running,
\beq
M_{\rm uv}\frac{d}{dM_{\rm uv}}\,\frac{\kappa}{g^{2}}=0\,.
\label{nonren}
\eeq
This means that the curvature ${\cal H}_{ik\bar j}$ is the renormalization group invariant tensor with no higher-loop corrections.
For a detailed derivation of the nonrenormalization theorem see Ref.\cite{Cui:2011rz}.

The situation is more complicated for
the ``main" coupling  $1/g^2$ appearing in the target space metric. The coupling in the bare Lagrangian
(i.e.\ at $M_{\rm uv}$) is holomorphic. This means that it can receive only one-loop renormalization, implying one-loop $\beta$ function.
Much in the same way as in four-dimensional ${\mathcal N}\! =\!1$ Yang-Mills theory
the holomorphic anomaly showing up in loops defies this theorem. The coupling constant the running of which is calculated in conventional perturbation theory is nonholomorphic. For the time being let us denote it by square brackets $1/[g^2]$, as in \cite{shiva}. In ${\mathcal N} =(2,2)$ sigma models
there is no holomorphy violation, and $1/g^2$ and $1/[g^2]$ coincide.

In other words,  in the undeformed (2,2) theory, i.e. at $\kappa=0$,
the holomorphicity of $1/g^{2}$ is maintained. It implies that only one-loop running
of $1/g^{2}$ is allowed, higher loops are absent. In the 4D case this phenomenon
 is also known in \mbox{${\mathcal N}\!=\!2$} gauge theories.
With less supersymmetry, i.e.\ in ${\mathcal N}\!=\!1$ gauge theories in 4D, holomorphicity is broken.
It happens usually at two-loop level but in certain cases appears already at the level of the first loop,
see Refs.\cite{Kapl,SVhola}. Likewise, our
 $\kappa$ term leads to breaking of holomorphicity
for $1/g^{2}$. This happens at the level of the first loop. More specifically, the first loop provides
a finite $|\kappa|^{2}$ correction to
$1/g^{2}$ which then leads to the nonholomorphic running of $g^2$ in the second loop.

\begin{figure}[h]
\begin{center}
\includegraphics[width=2in]{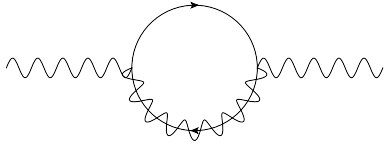}
\end{center}
\caption{\small One-loop finite correction to the canonical coupling $g$. The wave line denotes the background field $A$. The solid line denotes the propagator of $B$, while the solid line with a wavy line superposed denotes that of $\mathcal{B}$. We shall follow the same notation throughout this paper.}
\label{XXX}
\end{figure}

Iteration in $\kappa$ involves integration over the quantum $\zeta_{R}$ and $\psi_{R}$ fields
in the form of a polarization operator, see Fig.\,\ref{XXX}.
The polarization operator $\Pi_{RR}$ is defined as
\beq
\Pi_{RR}^{i\bar j}(x,y)=i |\kappa|^{2} \Big \langle T\left\{\zeta_{R}(x) \psi_{R}^{i}(x) \,\psi_{R}^{\dagger \bar j}(y)\zeta_{R}^{\dagger}(y)\right\} \Big\rangle_{\rm bck}
=i |\kappa|^{2}S_{\zeta}\,S^{i\bar j}_{\psi_{R}}\,,
\label{223}
\eeq
where $S_{\zeta}$ and $S_{\psi_{R}}$ are the propagators of $\zeta$ and $\psi_{R}$ fields in the background
of the bosonic field $\phi$. Referring to the Appendix B for details of calculation we give here the result for
$\Pi_{RR}$\,,
\beq
\Pi_{RR}^{i\bar j}(x,y)=-\frac{|\kappa|^{2}}{4\pi Z{\mathcal Z}}\,\langle x|\, G^{i\bar j} \,\nabla_{R }\,\frac{1}{\nabla_{L}\nabla_{R}}\,\nabla_{R}\,|y\rangle\,\,.
\label{prr}
\eeq

Let us emphasize that there is no ambiguity in the chiral fermion loop for $\Pi_{RR}$
due to its nonzero Lorentz spin. Generally speaking, the polarization operator $\Pi_{\mu\nu}$
can contain local terms such as $g_{\mu\nu}$ or $\epsilon_{\mu\nu}$\,.
These terms have zero Lorentz spin and do not contribute to $\Pi_{RR}$\,. Note also that in Eq.\,(\ref{prr}) the ordering
of operators is not important because we
neglect by commutator $\left[\nabla_{R},\nabla_{L}\right]^{i}_{k}\!=\!R_{k~m\bar n}^{~\,i}\!\left(\partial_{R}\phi^{\dagger\bar n}\partial_{L}\phi^{m}
\!-\!\partial_{L}\phi^{\dagger\bar n}\partial_{R}\phi^{m}\right)$. Additional terms with this commutator are infrared ones and do not contribute
to the running of couplings we are after.

In Appendix B we also explore an alternative derivation through a relevant UV regularization
via modification of the propagator for the $\zeta$ fermion. The result for $\Pi_{RR}$ is the same.

The fermion loop of Fig.\,\ref{XXX} then results in the following addition to the action,
\beq
\Delta_{\kappa}S\!=\!\!\int \! d^{2}x d^{2}y \,G_{i\bar k}\partial_{L} \phi^{\dagger \bar k}(x)\, \Pi^{i\bar j}i(x,y) \,\partial_{L}\phi^{l}(y) G_{l\bar j}
=\!-\frac{|\kappa|^{2}}{4\pi Z{\mathcal Z}}\!\int \!\! d^{2}x\, G_{i\bar j}\partial_{L}\phi^{\dagger \bar j} \partial_{R}\phi^{i}\,.
\eeq
Here we used  the relation
\beq
\nabla_{\!R}\,\partial_{L} \phi= \nabla_{\! L}\,\partial_{R} \phi\,,
\eeq
which makes the expression for the heterotic correction $\Delta_{\kappa} {\mathcal L}$ to the original bosonic Lagrangian local,
\beq
\Delta_{\kappa} {\mathcal L}=-\frac{|\kappa|^{2}}{4\pi Z{\mathcal Z}}\,G_{i\bar j}\partial_{R}\phi^{\dagger \bar j} \partial_{L}\phi^{i}\,.
\label{corL}
\eeq

The resulting correction to the metric $\Delta G_{i\bar j}$ can be rewritten in a more geometrical form in terms of the curvature ${\cal H}_{ik\bar j}$,
\beq
\Delta_{\kappa}G_{i\bar j}=-\frac{|\kappa|^{2}}{4\pi Z{\mathcal Z}}\,G_{i\bar j}=-\frac{1}{2\pi}\,{\cal H}_{\,lk\bar j}\xbar{\!\cal H}_{\bar l \bar k i}G^{(B)l\bar l} G^{(B)k\bar k}
=-\frac{1}{2\pi}\,{\cal H}_{\,lk\bar j}\xbar{\!\cal H}^{\,lk}_{\,i}\,,
\label{Gk}
\eeq
where the indices in $\xbar{\!\cal H}={\cal H}^{*}$ are raised by the $G^{(B)i\bar j}$ metric tensor (inverse to
$G^{(B)}_{i\bar j}$ defined in (\ref{GB})).

Equation (\ref{corL}) clearly demonstrates the breaking of holomorphicity by the fer-mion loop depicted in Fig.\,\ref{XXX}.
This loop is also related to the axial anomaly in the fermionic current. Indeed, as demonstrated in Appendix B,
while classically $\nabla_{L }\big( \zeta_{R} \psi_{R}^{i}\big)=0$,\footnote{\,Strictly speaking this divergence is not vanishing classically but additional terms do not contribute to $\Pi_{RR}$.}
 for the regularized loop of $\Pi_{RR}$ we get
\beq
\nabla_{L}\Pi^{i\bar j}_{RR}= -\frac{|\kappa|^{2}}{4\pi Z{\mathcal Z}}\,\langle x| \,\nabla_{R }\, G^{i\bar j}\,|y\rangle\,\,.
\label{UVk}
\eeq
Correspondingly we claim
the absence of higher-loop corrections to Eq.\,(\ref{corL}).

From the above considerations we see that perturbation theory is governed by two real couplings, $[g^{2}]$ and a real nonholomorphic
combination\,\footnote{\,The definition of $h^{2}$ in this paper corresponds to $\gamma^{2} g^4$ in \cite{Shifman:2008wv}. The reason for
this rescaling of the deformation parameter  compared to \cite{Shifman:2008wv} is that $g^2$ and $h^2$
 as defined here are the genuine loop expansion parameters.}
\beq
h^{2}=\frac{|\kappa|^{2}}{Z{\mathcal Z}}\,.
\label{227}
\eeq
We will also use the ratio $\rho$ of the couplings,
\beq
\rho\equiv \frac{h^{2}}{g^{2}}\,.
\label{227r}
\eeq

Then
Eq.\,(\ref{corL}) implies
\beq
\frac{1}{g^2} -\frac{1}{4\pi}\, [\rho] =\frac{1}{[g^2]}\,.
\label{226}
\eeq
It is convenient to rewrite Eq. (\ref{226}) as
\beq
\frac{1}{g^2}  =\frac{1}{[g^2]} + \frac{1}{4\pi} [\rho]\,,
\label{228}
\eeq
where the holomorphic coupling (i.e. renormalized only at one loop) on the  left-hand side is presented as a combination of two nonholomorphic terms.

\section{Beta functions}
\label{befu}
\setcounter{equation}{0}

\subsection{Generalities}
\label{gener}
Considering two couplings, $[g^{2}]$ and $h^{2}$, introduced above,
as functions of the UV cutoff $M_{\rm uv}$
we define two $\beta$ functions:
\beq
\beta_g \equiv \frac{d \,[g^2](M_{\rm uv})}{d L}\,,\qquad \beta_{h} \equiv \frac{d \,{ h}^2(M_{\rm uv})}{d L}\,,
\qquad L=\log\,M_{\rm uv}\,.
\label{betas}
\eeq
In what follows we will use also the $\beta$ function for $\rho$, see Eq.\,(\ref{227r}),
\beq
\beta_\rho \equiv \frac{d \,[\rho](M_{\rm uv})}{d L}\,.
\label{betar}
\eeq
We will omit below the square brackets in $[g^{2}]$ dealing with the 1PI definition of couplings.

As was discussed above nonrenormalization of the superpotential in Eq.\,(\ref{sfLa}) implies that the ratio $\kappa/g^{2}$ does not depend on $M_{\rm uv}$\,, see Eq.\,(\ref{nonren}). This equation can be rewritten as
\beq
\frac{d}{dL}\,\frac{|\kappa|^{2}}{g^{4}}=
 \frac{d}{dL}\,\frac{h^{2}Z{\mathcal Z}}{g^{4}}=\frac{h^{2}Z{\mathcal Z}}{g^{4}}\left[\frac{\beta_{h}}{h^{2}}-2\,\frac{\beta_{g}}{g^{2}}-\gamma\right]=0\,,
 \label{ghr}
\eeq
where the anomalous dimension $\gamma$ is defined as
\beq
 \gamma \equiv -\frac{d \,\log\, (Z{\mathcal Z})}{d L}=\gamma_{\psi_{R}}+\gamma_{\zeta}\,,\quad
 \gamma_{\psi_{R}} \equiv -\frac{d \,\log\, Z}{d L}\,,\quad
 \gamma_{\zeta} \equiv -\frac{d \,\log\, {\mathcal Z}}{d L}\,.
\label{andim}
\eeq

This fixes the $\beta$ function for ${h}^{2}$ in terms of the $\beta$ function for $g^{2}$
and the sum of the anomalous dimensions for $B^{i}$ and ${\mathcal B}$ fields,
\beq
\beta_{h}={h}^{2}\left[ \frac{2}{g^{2}}\,\beta_{g}+\gamma\right]\,.
\label{hg}
\eeq
For $\beta_\rho$ we get
\beq
\beta_{\rho}=\rho\left[ \frac{1}{g^{2}}\,\beta_{g}+\gamma\right]\,.
\label{rho_g}
\eeq

\subsection{Beta functions at one loop}

The relations (\ref{hg}) and (\ref{rho_g}) are exact to all loops. At the one-loop level all $\beta$ functions and anomalous dimensions
have been calculated earlier\,\cite{Cui:2011rz,Cui:2010si}:
\beqn
&&\beta_{g}^{(1)}=-T_{G}\,\frac{g^{4}}{4\pi}\,,\quad \gamma_{\zeta}^{(1)}=d\,\frac{ {h}^{2}}{2\pi}\,,\quad
\gamma_{\psi_{R}}^{(1)}=\frac{{ h}^{2}}{2\pi}\,,\quad \gamma^{(1)}=(d+1)\,\frac{{h}^{2}}{2\pi}\,,
\label{38a}\\[2mm]
&&
\beta_{h}^{(1)} =  - \frac{{h}^{2}}{2\pi }\left[ T_{G}g^2 - (d+1)h^{2}\right],\quad
\beta_\rho^{(1)} =  (d+1)\,\frac{h^{2}}{2\pi }\left[ \rho - \frac{T_{G}}{2(d+1)} \right].
\label{38b}
\eeqn
The results are  for the K\"ahler manifolds of the complex dimension $d$, which are homogeneous spaces $G/H$,
and $T_{G}$ is a dual Coxeter number of the group $G$. It is a straightforward generalization of calculations of Refs.\cite{Cui:2011rz,Cui:2010si},
where the CP($N\!-\!1$) sigma model was considered, for which $d=N - 1$ and $T_{G}=T_{{\rm SU}(N)}=N$.

An interesting feature of these one-loop results is that they
exhibit  a fixed point at $\rho=\rho_{c}=T_{G}/2(d+1)$, which becomes $\rho_{c}=1/2$ for CP($N\!-\!1$). At this point
\beq
\frac{\beta_{g}^{(1)} }{g^{2}}\Big|_{\rho=\rho_{c}}=\frac{\beta_{h}^{(1)} }{h^{2}}\Big|_{\rho=\rho_{c}}=-\,\gamma^{(1)} \Big|_{\rho=\rho_{c}}=-\frac{T_{G}g^{2}}{4\pi}\,,
\qquad \rho_{c}=\frac{T_{G}}{2(d+1)}\,.
\eeq

In terms of the geometrical interpretation we can present all one-loop results as corrections to the bosonic metric
$G_{i\bar j}$, and to the right-moving fermion extended metric $G^{(B)}_{i\bar j}$. These one-loop corrections are
\beq
\begin{split}
\Delta G_{i\bar j}\big|^{\rm one-loop}=-\frac{1}{2\pi}\left\{{\cal H}_{\,lk\bar j}\xbar{\!\cal H}_{\bar l \bar k i}G^{(B)l\bar l} G^{(B)k\bar k} +R_{i\bar j}\log M_{\rm uv}\right\},\\[2mm]
\Delta G^{(B)}_{i\bar j}\big|^{\rm one-loop}=-\frac{\log M_{\rm uv}}{2\pi}\left\{4{\cal H}_{\,ik\bar l}\xbar{\!\cal H}_{\,\bar j \bar k l}G^{(B)k\bar k}
G^{(B) l\bar l}+R^{(B)}_{i\bar j}\right\}.
\end{split}
\label{1loopL}
\eeq
As mentioned above there are no loop corrections
of any order to the heterotic curvature tensor ${\cal H}_{ik \bar j }$.

Let us parenthetically note that the parameter $(\kappa/g^{2})$ is related to $\delta$ introduced in \cite{Shifman:2008wv,Shifman:2008kj,B1},
where the large $N$ solution for  CP($N\!-\!1$) was constructed. Namely,
\beq
\frac{\kappa}{g} = \delta\,,
\eeq
(see the erratum to \cite{B1}). The $\delta$ parameter appears as the coefficient in a superpotential, see Eq.\,(C5) in \cite{B1}, and, as such, is also complexified.

In \cite{B1} it is shown that the physical parameter
determining (0,2) deformation is
\beq
u =\frac{16\pi}{N}\,\frac{\delta^2}{g^2}= \frac{16\pi}{Ng^2} \frac{\kappa^2}{g^2}\,,
\eeq
implying that (a) $u$ is proportional to $\kappa^2/g^4$ and, hence, is renormalization group invariant, as was expected; (b)
at large $N$ the physical parameter $u$ scales as $N^0$ while $\kappa^2$ and $g^2$ both scale as $1/N$.
The anomalous dimension $\gamma$ (see Eq.\,(\ref{andim})) then scales as $O(N^0)$, and $\beta_g$ becomes one-loop exact in the limit $N\to\infty$, see Eq.\,(\ref{11}).

\section{Beyond one loop from instanton calculus}
\setcounter{equation}{0}

In this section we will briefly outline the instanton derivation of the $\beta$ functions along the lines
of \cite{Novikov,Cui:2011rz,Cui:2011uw,shiva}. In particular, we will use the nonrenormalization theorem for the second and higher loops in the instanton background. Only zero modes and the one-loop contribution have to be considered.
All parameters that will appear in the derivation below are those from the bare Lagrangian.

To warm up let us briefly review the instanton calculation in \cite{Novikov}.
Consider the instanton measure in four-dimensional ${\mathcal N}=1$ Yang-Mills theory with the SU$(N)$ gauge group.
In the quasiclassical approximation the renormalization group invariant (RGI) prefactor in the instanton measure is
\beq
\mu_{\rm inst}^{(1)} = M_{\rm uv}^{3N} \exp\left(-\frac{8\pi^2}{g^2}\right)\,.
\label{im1}
\eeq
The factor in the exponent is the classical instanton action, while the pre-exponential factor
comes from the zero modes. There are $n_{B}=4N$ bosonic zero modes and $n_{F}=2N$ fermion ones
to produce $M^{n_{B}}/M^{n_{F}/2}=M^{3N}$. In terms of perturbation theory
Eq.\,(\ref{im1}) gives us the one-loop running of the holomorphic coupling.
What happens in higher loops?

The only change (to all orders in the coupling constant) is the emergence of another pre-exponential factor $g^{\,n_{F}}/g^{\,n_{B}}=1/g^{\,2N}$, due to
normalization of the zero modes, namely,
\beq
\mu_{\rm inst}^{\rm exact} = M_{\rm uv}^{3N}\, \frac{1}{[g^2]^N} \, \exp\left(-\frac{8\pi^2}{[g^2]}\right)\,.
\label{im2}
\eeq
Simultaneously, $ {1}/{[g^2]}$ in the exponent becomes nonholomorphic. The combination (\ref{im2}) of $M_{\rm uv}$ and
$[g^2](M_{\rm uv})$ is renormalization group invariant.
Differentiating over $\log M_{\rm uv}$ we arrive at the NSVZ $\beta $ function for SU($N$). Generalization to an arbitrary
gauge group $G$ is just a substitution of $N$ in expressions above by the dual Coxeter number $T_{G}$.

If matter fields are added, the only further changes in $\mu_{\rm inst}^{\rm exact}$ are as follows: (i) the power of $M_{\rm uv}$ is changed appropriately and (ii) $Z^{-1/2} $ factor appears in the pre-exponent for each matter-sector fermion zero mode. The number of such fermion zero modes
is given by $2T(R)$. where $T(R)$ is the Dynkin index of representation $R$.
 In this way one obtains the full exact NSVZ $\beta$ function,
\begin{equation}
\beta_{\,\rm NSVZ} (g^2) = -\frac{g^4}{8\pi^2}\left[3\,T_G -\sum_{\rm matter} T(R_i)(1-\gamma_i )
\right]\left(1-\frac{T_G\,g^2}{8\pi^2} \right)^{-1}
\, .
\label{nsvzbetaf}
\end{equation}

\vspace{2mm}

Now, let us see how the same strategy can be implemented in the $\kappa$ deformed  CP$(N\!-\!1)$ sigma model under consideration.
Let us start first with the nondeformed  (2,2) case. At the classical level the instanton-generated exponent is
\beq
\exp\left( -\frac{4\pi}{g^2}\right)\,,
\label{415}
\eeq
see e.g. \cite{Novikov2}. At the one-loop level (and at higher loops as well) nonzero modes cancel out.
In the CP$(N-1)$ model  there are $n_{B}=2N$ bosonic zero modes and $n_{F}=2N$ fermion zero modes,
This produces the $M^{n_{B}}/M^{n_{F}/2}=M^{N}$  pre-exponential factor. As for normalization of the zero modes
the corresponding factors cancel out between bosonic and fermion modes, $g^{n_{B}}/g^{n_{F}}=1$.
Thus, we come to
\beq
\mu_{\rm inst}(2,2)=M^{N}\exp\left(-\frac{4\pi}{g^{2}}\right)={\rm RGI}\,,
\eeq
which leads to the one-loop exact $\beta$ function and unbroken holomorphicity in the (2,2) theory.

Now let us switch on the $\kappa$ modification. As we discussed in Sec.\,\ref{sec:holo} the holomorphicity is broken
in perturbation theory already in the order $|\kappa|^{2}$. Here comes a surprise: such breaking does not occur
in the instanton background\,!

Indeed, the $\kappa$ terms in the Lagrangian (\ref{components}) has the form
\beq
i\kappa\,G_{i\bar j}\, \zeta_R \,\psi_R^{i}\,\partial_{L}\phi^{\dagger\bar j} -i \kappa^{*}G_{i\bar j} \,\psi_{R}^{\bar j}\,\zeta_{R}\,\partial_{L}\phi^{i}\,;
\eeq
the product of these two terms enters in the fermion loop calculation. After Euclidean continuation the instanton (or anti-instanton) background
leads to vanishing either  $\partial_{L}\phi^{\dagger\bar j}$ or  $\partial_{L}\phi^{i}$. Therefore,  the  $|\kappa|^{2}$ iteration is not possible.
It means that holomorphicity is not broken in one loop for the instanton, and the instanton action stays $4\pi/g^{2}$ with the original $1/g^{2}$.
In terms of the running $1/[g^{2}]$ and $[\rho]$ it means that the combination (\ref{228}) enters into the
instanton exponent,
\beq
\exp\left( -\frac{4\pi}{g^2}\right)=\exp\left(-\frac{4\pi}{[g^2]}-[\rho]\right).
\eeq

Moreover, in the instanton background the only effect of an additional right-mover $\zeta_{R}$ is its admixture to $\psi_{R}$.
This triangle mixing does not change the eigenvlues so the cancellation of nonzero modes stays the same as in the (2,2) case.
Also, the counting of the zero modes does not change. What appears in the pre-exponential factor is an additional $Z^{-1/2}$ factor
for each zero mode of $\psi_{R}$ because of its normalization \cite{JC}. There are $N$ such modes so we arrive at
 the following exact expression for the measure:
\beq
\mu_{\rm inst}^{\rm exact} = \frac{1}{Z^{N/2}}\,M_{\rm uv}^{N}\,
\exp\left(-\frac{4\pi}{[g^2]}-[\rho]\right)\,.
\label{im4}
\eeq
The $Z$ factor is defined in (\ref{sfLa}), see the first term in the second line.
 All effects due to two loops and higher, associated with nonzero modes,
cancel \cite{Cui:2011rz,Cui:2011uw} much in the same way as in the (2,2) case.

Equation (\ref{im4}) implies
\beq
 T_{G} \log M_{\rm uv} -  \frac{T_{G}}{2} \log Z-\frac{4\pi}{[g^2]}-[\rho]= {\rm RGI} \,,
\eeq
where we substitute $N$ by the Coxeter index to consider a generic K\"ahler manifold $G/H$.

Differentiating over $\log M_{\rm uv}$ and using Eqs.\,(\ref{andim}) and (\ref{rho_g}) for $\beta_\rho$ we arrive at
the full exact $\beta $ function (\ref{11}), relating $\beta_g$ to the anomalous dimensions, much in the same way as the NSVZ formula.

Combining the full $\beta$ function in (\ref{11}) with Eqs.\,(\ref{hg}) and (\ref{rho_g})
we derive the ``secondary" $\beta$ functions,
\beq
\begin{split}
&\beta_h = -\frac{h^2}{1-(h^{2/}4\pi)}\left[T_{G}\,\frac{g^2}{2\pi}\left(1+\frac 12\gamma_{\psi_R}\right) -\gamma\left(1 +\frac{h^2}{4\pi}\right)\right],
\\[4mm]
&\beta_\rho = -\frac{\rho}{1-(h^{2}/4\pi)}\left[ T_{G}\,\frac{g^2}{4\pi}\left(1+\frac 12\gamma_{\psi_R}\right) -\gamma\,\right].
\end{split}
\label{berh}
\eeq

\section{Explicit two-loop calculations}
\label{etlc}
\setcounter{equation}{0}

\subsection{Two-loop {\boldmath $\beta$} function for {\boldmath $g^2$}}
\label{sec:superfg}

In this section we will use the superfield method to calculate two-loop beta function for $g^2$ for the heterotic model at any symmetric K\"ahler target space. We will use a linear background field method, setting the background field $A_{bk}=fe^{-i x\cdot k}$ (see the review paper \cite{Novikov2}). The basic method is roughly the same as that of component field. We expand the action around the chosen background, and calculate all relevant diagrams.
To maintain supersymmetry, we use supersymmetric dimensional reduction, which in turn reduces to dimensional regularization in our case. Note that since we are only interested in the renormalization of the canonical coupling, this is compatible. Also we will keep the vector current of the theory conserved. Due to the computational nature, it would not be beneficial to show all steps in detail here. Instead, we will offer some intuitive arguments for the reader to understand our results. For a detailed description of the calculational method and examples, the reader is referred to \cite{Cui:thesis}.

To keep our discussion concise, we will only show those Feynman diagrams that are of the leading order with respect to target space curvature (i.e. assuming $\phi$ and $\phi^\dagger$ to be small).

\begin{figure}[h]
\begin{center}
\includegraphics[width=4.4in]{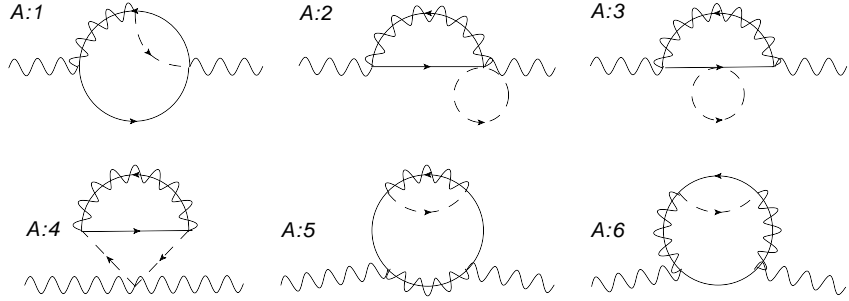}
\end{center}
\caption{\small Two-loop correction to the canonical coupling $g$. The dashed line denotes the quantum propagator of $A$.}
\label{h2loopg}
\end{figure}

At two-loop level, the correction of the order $g^4$ is obviously absent, as predicted by the undeformed model. For the correction of the order $g^2 h^2$ and $h^4$, the relevant diagrams are those  shown in Fig.\,\ref{h2loopg} and Table\,\ref{g2looppoles}, at leading order (with respect to the covariant structure) contributed by the superfield $A$, which renormalizes $g^2$.

\begin{table}
\begin{center}
\begin{tabular}{| c || c | c |}
\hline
Diagram & Double pole & Single pole \rule{0mm}{6mm}\\[2mm] \hline
A:1 &  $0$ & $-T_{G}I{g_{0}^2h_{0}^2}/{4\pi}$\rule{0mm}{6mm}\\[2mm]
A:2 &  $0$ & $T_{G}I{g_{0}^2h_{0}^2}/{4\pi}$ \rule{0mm}{6mm}\\[2mm]
A:3 & $0$ & $-({T_{G}}/2)I{g_{0}^2h_{0}^2}/{4\pi}$\rule{0mm}{6mm}\\[2mm]
A:4 & $0$ & $-({T_{G}}/2)I{g_{0}^2h_{0}^2}/{4\pi}$ \rule{0mm}{6mm}\\[2mm]
A:5 & $0$ & $I{h_{0}^4}/{4\pi}$\rule{0mm}{6mm}\\[2mm]
A:6 & $0$ & $d\,I{h_{0}^4}/{4\pi}$\rule{0mm}{6mm}\\[2mm]
\hline
\end{tabular}
\end{center}
\caption{\small Two-loop calculation for the $g^2$ correction. The labeling of the diagrams  follows that in
Fig.\,\ref{h2loopg}.}
\label{g2looppoles}
\end{table}

Now it is rather straightforward to show that these diagrams, together with the Hermitian conjugated part, give rise to the following expression:
\beq
\begin{split}
\frac 1{g^2(\mu)} = &~~~\,\frac 1{g_0^2}\left[1-\frac {h_{0}^2}{4\pi}-\frac{T_{G}}{2} \,g_{0}^{2} I\right] \\[2mm]
&+ \frac 1{g_0^2}\left[-\frac{T_{G}}{4\pi}\, g_{0}^{2}h^2 I+\frac{d+1}{4\pi}\,h_{0}^4\, I\right],
 \end{split}
\label{41}
\eeq
where
\beq
I=\frac {1}{2\pi}\, {\rm log} \left(\frac{M_{\rm uv}}{\mu}\right)\,.
\label{42}
\eeq
For a more accurate definition of $I$ in (\ref{42}) in terms of a dimensionally regularized loop integral
see Refs.\cite{Cui:2010si,Cui:thesis}.

The first line in (\ref{41}) contains the one-loop contributions  and the second one is the result for two-loop diagrams.
From (\ref{41}) we get the two-loop $\beta$ function,
\beq
\beta_{g}^{(2)}= -\frac{g^2}{4\pi}\left[T_{G}g^{2}\Big(1+\frac{h^2}{2\pi}\Big)-(d+1)\,\frac{h^4}{2\pi}\right] \,.
\label{2loopgform}
\eeq
This coincides with the corresponding expansion of the master expression (\ref{11}). The one-loop expressions (\ref{38a}) for
the anomalous dimensions are sufficient
for this comparison.

\subsection{Anomalous dimensions at two loops}

\begin{figure}[h]
\begin{center}
\includegraphics[width=4in]{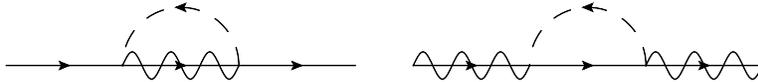}
\end{center}
\caption{\small One-loop correction to the wave-function renormalization of $\psi_R$ and $\zeta_R$.}
\label{1loopgamma2}
\end{figure}

Now we turn to the calculation of the $\beta$ function of the deformation coupling $h$. To this end we will have to understand anomalous dimensions of the fermionic fields $\psi_R$ and $\zeta_R$ first. At one-loop level they are given in Eq.\,(\ref{38a}), see Fig.\,\ref{1loopgamma2} for the corresponding diagrams.

\begin{figure}[h]
\begin{center}
\includegraphics[width=5.5in]{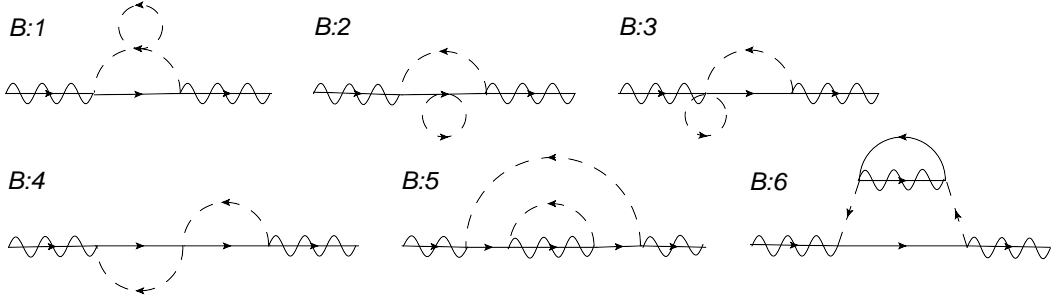}
\end{center}
\caption{\small Two-loop corrections in the wave-function renormalization of $\zeta_R$\,. }
\label{h2loopzeta}
\end{figure}

\begin{figure}[h]
\begin{center}
\includegraphics[width=5.5in]{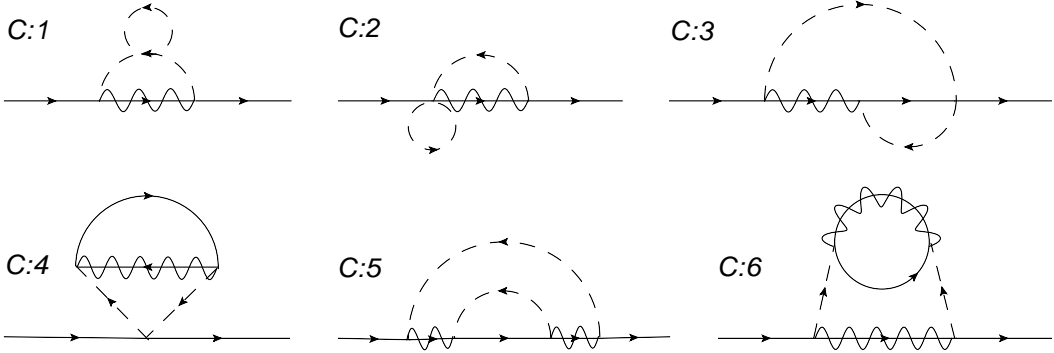}
\end{center}
\caption{\small Two-loop  corrections to the wave-function renormalization of $\psi_R$\,.}
\label{h2looppsi}
\end{figure}

At two-loop level we know that at the order $g^4$ there is no correction. At the orders $g^2h^2$ and $h^4$ we have the  diagrams (in superfields) shown in Fig.~\ref{h2loopzeta} and Fig.~\ref{h2looppsi} that contribute to $\gamma_{\zeta}$  and $\gamma_{\psi_R}$, respectively.

The renormalization of $\zeta_{R}$ is easier to understand as ${\cal Z}$ is obtained by evaluating all  diagrams in Fig.\,\ref{h2loopzeta}. Assembling them all we have
\beq
{\cal Z} = 1+d\,h_{0}^2 I+d\,\frac{h_{0}^4}{4\pi}\,I +d\,T_{G}\,\frac{h_{0}^2g_{0}^2}2I^2-d \,\frac{h_{0}^4}{2} I^2\,.
\\
\rule{0mm}{3mm}
\eeq
The two-loop anomalous dimension $\gamma_{\zeta}$  then can be written as
\beq
\gamma_{\zeta}^{(2)}=-\frac{1}{\cal Z}\,\mu\frac{d{\cal Z}}{d\mu}=
d\,\frac{h_{0}^{2}}{2\pi}\left(1+\frac{h_{0}^{2}}{4\pi}\right)\left[1+I\big(T_{G}g_{0}^{2}-(d+1)h_{0}^{2}\big) \right]\,.
\eeq
The second factor in the rhs (the square brackets) just shifts  $h_{0}^{2}$ to $h^{2}(\mu)$ in accord with the one-loop
$\beta_{h}$ given in Eq.\,(\ref{38b}).  Thus, we get
\beq
\gamma_{\zeta}^{(2)}=
d\,\frac{h^{2}}{2\pi}\left(1+\frac{h^{2}}{4\pi}\right).
\label{2loopz}
\eeq

In the case of the wave-function renormalization of $\psi_{R}$
it should be noted that the diagrams shown in
Fig.\,\ref{h2looppsi} in fact do not directly contribute to $Z$, but, rather, to $Z/g^2$. Therefore,  we have
\beq
\frac{Z}{g^2} = \frac{1}{g_{0}^{2}}\left[1-T_{G}\,\frac{g_{0}^2}2 \,I+h_{0}^2 I+\frac{h_{0}^4}{4\pi}\, I - d\,\frac{h_{0}^4}2 I^2 -T_{G}\, \frac{h_{0}^2g_{0}^2}{8\pi} I\right].
\label{45}
\eeq
Using Eq.\,(\ref{42}) for $1/g^{2}$ we get for $Z$,
\beq
Z=1+h_{0}^2 I +T_{G}\, \frac{h_{0}^2g_{0}^2}{8\pi}\,I-d\,\frac{h_{0}^4}{4\pi}\, I +T_{G}\frac{g_{0}^{2}h_{0}^{2}}{2}\, I^{2} - d\,\frac{h_{0}^4}2 I^2 \,.
\eeq
It leads to the following  two-loop anomalous dimension of $\psi_{R}$\,,
\beq
\gamma_{\psi_{R}}^{(2)}=-\frac{1}{ Z}\,\mu\frac{d{Z}}{d\mu}=
\frac{h_{0}^{2}}{2\pi}\Big(1+T_{G}\, \frac{g_{0}^2}{8\pi}-d\,\frac{h_{0}^2}{4\pi}\Big)\left[1+I\big(T_{G}g_{0}^{2}-(d+1)h_{0}^{2}\big) \right]
\eeq
Again, the factor in the square brackets containing $I$  just shifts $h_{0}^{2}$ to $h^{2}(\mu)$, so
\beq
\gamma_{\psi_{R}}^{(2)}=
\frac{h^{2}}{2\pi}\Big(1+T_{G}\, \frac{g^2}{8\pi}-d\,\frac{h^2}{4\pi}\Big).
\label{2looppsi}
\eeq

\subsection{Beta functions and  fixed point in {\boldmath $\rho$}}
\label{sec:fixed}
Knowledge of two-loop anomalous dimensions means that we know $\beta_{g}$
at three-loop level. The explicit expression for $\beta^{(3)}_{g}$ follows from substitution of the anomalous dimensions
(\ref{2loopz}) and (\ref{2looppsi}) into the master formula (\ref{11}),
\beq
\beta^{(3)}_{g}=-\frac{g^{2}/4\pi}{1-(h^{2/}4\pi)}\left[T_{G}\,g^{2}+\frac{h^{2}}{4\pi}\Big(T_{G}g^{2}-2(d+1)h^{2}\Big)\left(1+
T_{G}\,\frac{g^{2}}{8\pi}\right)\right].
\eeq

For $\beta_{h}$ and $\beta_{\rho}$ we get the two-loop expressions,
\beqn
&&\beta^{(2)}_{h}=-\frac{h^{2}/2\pi}{1-(h^{2}/4\pi)}\left[T_{G}g^{2}-(d+1)h^{2}+\frac{h^{2}}{8\pi}\Big(T_{G}g^{2}-2(d+1)h^{2}\Big)\right],
\\[3mm]
&&\beta^{(2)}_{\rho}=(d+1)\,\frac{g^{2}}{2\pi}\,\frac{\rho}{1-(h^{2}/4\pi)}\Big(\rho -\frac{T_{G}}{2(d+1)}\,\Big).
\eeqn
The expression for $\beta_{\rho}$ differs from the one-loop expression (\ref{38a}) only by a factor, so the fixed point $\rho_{c}=T_{G}/2(d+1)$ stays
intact. Certainly, it would be interesting to find a geometrical interpretation of this fixed point but we do not have an answer for this yet.

At this point
\beq
\beta^{(3)}_{g}\Big |_{\rho=\rho_{c}}\!=\!-T_{G}\,\frac{g^{4}}{4\pi}\,\frac{1}{1-(h^{2}/4\pi)}\,,\qquad \beta^{(2)}_{h}\Big |_{\rho=\rho_{c}}\!=\!-T_{G}\,\frac{h^{2}g^{2}}{4\pi}\,\frac{1}{1-(h^{2}/4\pi)}
\eeq
differ only by a factor $1/(1-(h^{2}/4\pi))$ from the corresponding one-loop results.

\section{Isometries of the model}
\label{isomcurr}
\setcounter{equation}{0}

In this section we study the isometries of the heterotic models and, in particular, address the question of whether
they could be broken by loop corrections.
For generic sigma model there are the following symmetry transformations of bosonic fields
$\phi^{\,i}$, $\phi^{\dagger \,\bar j}$ living on the K\"ahler target space:
\beq
\phi^{\,i}\to \phi^{\,i}+ \epsilon^{\,A} V_{A}^{i}(\phi)\,,\quad \phi^{\dagger \,\bar i}\to \phi^{\dagger \,\bar i}+\epsilon^{\,A} \xbar{V}_{A}^{\,\bar i}(\phi^{\dagger})\,,
\eeq
where the vector $V^i_A$ is the Killing vector over the target manifold, $\epsilon^A$ are real infinitesimal parameters, and the index $A$ labels   isometries. Note that in the K\"ahler cases, the Killing vector $V_A$ has only holomorphic dependence on the bosonic \mbox{field $\phi$.}

We are dealing with symmetric homogeneous spaces $G/H$\,. Correspondingly, isometries arising from the algebra of $H$ are
realized linearly, while the remaining generators in the algebra of the group $G$ are realized nonlinearly, these symmetries
are spontaneously broken. For example, in
${\rm CP}(N\!-\!1)={\rm SU}(N)/{\rm S}({\rm U}(N\!-\!1)\times {\rm U}(1))$, we have $(N\!-\!1)^2$ linear symmetries corresponding to ${\rm U}(N\!-\!1)$ rotations of fields $\phi^i\,,~ \phi^{\dagger\bar{j}}$. The remaining $2N-2$ symmetries are nonlinearly realized. They can be written as
\beq
\phi^i\to \phi^i+ \epsilon^{i\bar{j}}\phi_{\bar{j}}+\beta^{i}+(\beta^{\dagger}\phi)\phi^{i}\,,\quad \phi^{\dagger\bar{j}}\to \phi^{\dagger\bar{j}}- \epsilon^{i\bar{j}}\phi^{\dagger}_{i}+\beta^{\dagger\bar{j}}+(\beta\phi^{\dagger})\phi^{\dagger\bar{j}}\,,
\label{IsoT}
\eeq
where the indices of charts $\{ \phi^i, \phi^{\dagger\bar{j}}\}$ locally are raised or lowered by $\delta^{i\bar{j}}$ or $\delta_{i\bar{j}}$.

One can supersymmetrize the above model, and write down the general form in the $\mathcal{N}\!=\!(2,2)$ case. It can be done in terms of superfields by simply promoting $\phi^i$ and $\phi^{\dagger\bar{j}}$ to chiral and antichiral superfields. In components, the fermions $\psi^{i}$  living on tangent space of CP$(N\!-\!1)$ transform as tensors corresponding to isometries
\beq
\psi^i_{R,L}\to \psi^i_{R,L}+ \epsilon^{\alpha}\partial_j V_{A}^i(\phi)\psi^j_{R,L}\,.
\label{IsoTf}
\eeq
Turning on heterotic deformation does not change the isometries, the additional fermion field $\zeta_{R}$ is a singlet of the group $G$ action.
 These symmetries can be verified classically in the geometric formulation of the Lagrangian (\ref{Lb}), as long as the curvature ${\cal H}_{ik\bar j}$ satisfies:
\beq
{\cal L}_A{\cal H}=0\,,
\eeq
where ${\cal L}_A$ is the Lie derivative with respect to the $A$th isometry. In the heterotic case, the only nontrivial components of ${\cal H}_{ik\bar j}$ are proportional to the metric $G_{i\bar{j}}$, see Eq.\,(\ref{Hcomp}). It apparently satisfies the above condition. However,
the heterotic coupling leads to a change in the expression for the isometry current $J_{R}^{A}$ as compared with the (2,2) model,
\beq
J^{A}_{R}=\frac{1}{2}\,\xbar{V}_{A}^{\,\bar j}G_{i\bar j}\partial_{R}\phi^i+\frac{i}{2}\,\nabla_k V^{\,i}_{A}G_{i\bar j}\psi_{R}^{\dagger \bar{j}}\psi_R^k
+i\kappa\,\xbar{V}^{\,\bar j}_{A}G_{i\bar j}\zeta_R\psi_R^i+{\rm H.c.}\,,
\eeq
while $J^{A}_{L}$ does not change.

The question to ask is whether or not the deformation under consideration would deform the classical geometry, since now the chiral fermion $\zeta_{\,R}$ enters these currents.
To answer this, we need to see if these isometric currents have anomalies. It could be verified  either by calculating anomalies of these currents or by checking the isometry transformations of the effective action after the one-loop correction. We will proceed along the second route  because it is easier and more transparent to demonstrate the isometry invariance in the effective action.

Moreover, even in case when isometry currents happened to be anomalous it does not imply breaking of isometries, anomaly could be
a total derivative and does not lead to nonconservation  for corresponding generators, at least, in perturbation theory. For this reason examination
of the effective action is preferable.

It is worth noting that, if there is any anomaly, it happens due to the fermionic loops. Therefore, we can choose nonzero only  bosonic background, and consider the fermion loop corrections to the effective action. Up to one-loop order, keeping terms bilinear in fermionic fields is sufficient. As a result the relevant part of Lagrangian takes the form
\beq
\begin{split}
{\mathcal{L}}_{\rm ferm} &= {\cal Z}\,\zeta_R^\dagger\! \left(1+ \frac{\partial_{\mu}\partial^{\mu}}{M^{2}}\right)\! i\partial_L \, \zeta_R +G_{i\bar j}\!\left[\psi_{L}^{\dagger \bar j}\,i\nabla_{\!R}\,\psi_{L}^{i}
+Z \psi_{R}^{\dagger \bar j}\,i\nabla_{\!L}\psi_{R}^{i}\right]\\[2mm]
 &~~+ \left[\kappa\, \zeta_R  \,G_{i\bar j}\big( i\,\partial_{L}\phi^{\dagger \bar j}\big)\psi_R^{i}
+{\rm H.c.}\right],
\end{split}
\label{6.6}
\eeq
where all bosonic fields are background, while fermionic fields are to be integrated out. We also introduced here regularization for the $\zeta_{R}$
field by introducing higher derivatives. This regularization which makes loops with $\zeta_{R}$ convergent is clearly consistent with isometries.
It proves then that the heterotic modification of the (2,2) theory does not break any (2,2) isometry.

Our explicit one-loop calculation in Appendix B confirms this. In addition, we also present here geometry of the target space introducing
vielbeins $e^{a}_{\ i}$ and $\bar{e}^{\ \bar{b}}_{\bar{j}}$ to factorize
the
metric tensor $G_{i\bar{j}}$ and make sure our effective action preserves explicit geometric structure. Since the fermion fields naturally live on the tangent space, we also redefine the fermions $\psi$ to transform the Lagrangian to  the canonical form,
\beq
\begin{split}
&e^{\,a}_{\ i}\,e^{\,i}_{\ b}=\delta^{\,a}_{\ b}\,, \quad \bar{e}_{\bar{a}}^{\ \bar{j}}\,\bar{e}^{\ \bar{b}}_{\bar{j}}=\delta_{\bar{a}}^{\ \bar{b}}\,, \quad
\delta_{a\bar{b}}\,e^{\,a}_{\ i}\,\bar{e}^{\ \bar{b}}_{\bar{j}}=G_{i\bar{j}}\,;\\[2mm]
&~~~~~~~~\psi^{\,a}_{R,L}\equiv e^a_{\ i}\,\psi^i_{R,L}, \quad \bar{\psi}^{\,\bar{b}}_{R,L}\equiv \bar{\psi}^{\,\bar{j}}_{R,L}\,\bar{e}^{\ \bar{b}}_{\bar{j}}\,,
\end{split}
\eeq
where the tensor indices $\{i, \bar{j}\}$ of vielbein are lowered or raised by the metric $G_{i\bar{j}}$ and $G^{\bar{j}i}$, and the frame indices $\{a, \bar{b}\}$ by the flat metric $\delta_{\bar{b}a}$ and $\delta^{a\bar{b}}$. After these redefinitions the Lagrangian can be rewritten as
\beqn
{\mathcal{L}}_{\,\rm ferm}\!=\!{\cal Z} \zeta_R^\dagger \Big(1\! +\!  \frac{\partial_{\mu}\partial^{\mu}}{M^{2}}\Big) i\partial_L \zeta_R \! +\psi_{L a}^{\dagger}\,i\widetilde{\nabla}_{\!R}\,\psi_{L}^{a}
\! + Z\psi_{Ra}^{\dagger}i\widetilde{\nabla}_{\!L}\psi_{R}^{a} \!  +\! \left[i\kappa \,\zeta_R\, \bar{e}_{La}\,\psi_R^{a}
\! +\! {\rm H.c.}\right],
\label{Lferm}
\eeqn
where
$$\widetilde{\nabla}_{\!R,L}\,\psi_{L,R}^{a}=\partial_{R,L}\psi_{L,R}^{a}+\Omega^{\ \ a}_{R,L\,c}\,\psi_{L,R}^{c}$$
and $$\bar{e}_{La}=\bar{e}_{\bar{j}a}\partial_L\phi^{\dagger\bar{j}}\,.$$ Moreover,  $\Omega^{a}_{R,Lb}$ and $\bar{e}_{La}$ are pull-back spin-connection on the frame bundle and vielbeins, respectively. We express the fermion kinetic term canonically, and the isometries are realized in terms of the frame bundle indices $\{a, \bar{b}\}$ rather than $\{i, \bar{j}\}$.

Next we find the isometry transformations on fermions, vielbeins and spin-con\-nection. It is actually clear how the transformation must look like. Geometrically, once we perform the isometry transformation, there will be effectively an induced rotation on the frame bundle. Now fermions and vielbeins are matter type fields, while spin-connections are gauge fields with respect to U$(N\!-\!1)$ gauge symmetries. Therefore, the general form of isometry transformation is
\beq
\begin{split}
&\delta\psi^{\,a}_{R,L}=v^{\,a}_{A\ c}\,\psi^{\,c}_{R,L}\,,\quad \delta\bar{\psi}_{R,L\,a}=-\bar{\psi}_{R,L\,c}\, v^{\,c}_{A\ a}\,,\\[2mm]
&\delta e^{\,a}_{L}=v^{\,a}_{A\ c}\,e^{\,c}_{L}\,,\qquad ~~~\delta\bar{e}_{\!La}=-\bar{e}_{L\,c}v^{\,c}_{A\ a}\,,\\[2mm]
&\delta\Omega^{\ \ \ a}_{R,L\,c}=-\partial_{R,L} v^{\,a}_{A\ c}-\big[\Omega_{R,L}\,, v_A\big]^a_{\ c}\,.
\end{split}
\label{frT}
\eeq
The explicit expression of $v^a_{A\ c}$ can be found from Eqs.\,(\ref{IsoT}) and (\ref{IsoTf}), once the vielbeins are given. However,  the explicit form of $v^a_{\ c}$ is not significant, Eq.\,(\ref{frT}) is all we need.

Now, when isometries of the Lagrangian are verified and regularization provides convergence of fermion loop integration we can claim that all original target space isometries are preserved under the heterotic modification. Furthermore the transformation rule of isometries, Eq.\,(\ref{frT}), is a special case of general holonomy transformations on a frame bundle. Therefore the heterotic model is free of holonomy anomaly as well. At last we want to emphasize that we can always introduce appropriate regulators without breaking target space isometries, higher derivative for example, so long as the chiral fermion $\zeta_R$ couples to an isometry-invariant term, see Eq.\,(\ref{6.6}). It is essentially different from the situation that chiral fermions couple to gauge fields or spin-connections where gauge symmetries or target space symmetries can be only preserved conditionally \cite{Moore:1984ws, CCSV2}.

\section{Supercurrent multiplet}
\label{supercurr}
\setcounter{equation}{0}

In this section we analyze the hypercurrent, a superfield which contains supercurrent and energy-momentum tensor
among its components. For undeformed ${\mathcal N}\!=\!(2,2)$ theories the hypercurrent and its quantum anomalies
were studied in Ref.\cite{SVZ}. This study includes, in particular, the anomaly in the central charge which does not
enter the \nzt algebra.
The general formulation in case of $\mathcal{N}=(0,2)$ theories was given in Ref.\cite {Dumitrescu:2011iu},
see also the earlier references \cite{Witten:1993yc} and related considerations \cite{Benini:2012cz}.
We present an explicit superfield form for the hypercurrent and all anomalies in the heterotic models under consideration.

\subsection{Hypercurrent in the undeformed {\boldmath ${\mathcal N}\!=(2,2)$} theory }
Let us start with the definition of the hypercurrent ${\cal T}_{\mu}$ in the undeformed ${\mathcal N}=(2,2)$ theory.
The hypercurrent is the supermultiplet containing a supersymmetry current $s_{\mu\alpha}$ and an energy-momentum tensor
$\vartheta_{\mu\nu}$\,,
\begin{equation}
{\cal T}_{\mu}={v}_{\mu}+\big[ \theta \gamma^{0}s_{\mu} +{\rm H.c.}\big]
-2\,\bar\theta \gamma^{\nu}\theta\, \vartheta_{\mu\nu}
+\ldots \,.
\end{equation}
Here $\theta$ is the spinor $\theta\!=\!(\theta^{1},\theta^{2})\!=\!(\theta_{L}\,, \theta_{R})$ and the lowest component
${v}^\mu=G_{i\bar j}\,\bar\psi^{\bar j}\gamma^{\mu}\psi^{i}$ is the fermionic $R$ current.

 Introducing spinor indices,
${\cal T}_{\alpha\beta}=(\gamma^{0}\gamma^{\mu})_{\alpha\beta}{\cal T}_{\mu}$ we can write the classical hypercurrent in terms of the
${\mathcal N}=(2,2)$  chiral superfields $\Phi^{i}(x,\theta)$,
\beq
{\cal T}_{\beta\alpha}=G_{i\bar j}\bar D_{\beta} \Phi^{\dagger\,\bar j}D_{\alpha}\Phi^{i}\,,
\eeq
where $D_{\alpha}$, $\bar D_{\beta}$ are conventional spinor derivatives
and the metric $G$
is a function of superfields $\Phi^{i}\,,\Phi^{\dagger\,\bar j}$. Actually, only components ${\cal T}_{11}$ and
${\cal T}_{22}$, presenting nonzero Lorentz spin, are associated with the hypercurrent ${\cal T}_{\mu}$\,, the scalar
${\cal T}_{12}=[{\cal T}_{21}]^{\dagger}$ represents the twisted chiral integrand in the superspace action.

The anomaly equations for the hypercurrent derived in \cite{SVZ} are of the form:\footnote{\,There is a misprint
in \cite{SVZ}: the quantum anomalies of the hypercurrent should be multiplied by the overall factor $(-1/2)$\,.}
\begin{equation}
\label{supanom1p}
\begin{split}
\xbar{\!D}_{1}{\cal T}_{22}=\frac{1}{4\pi}\,\xbar{\!D}_{2} \,\Big[R_{i\bar j}\bar D_{1} \Phi^{\dagger\,\bar j}D_{2}\Phi^{i}\Big]=\frac{T_{G}g^{2}}{8\pi}\,\bar D_{2} \,{\cal T}_{12}
\,,\\[2mm]
\bar D_{2}{\cal T}_{11}=\frac{1}{4\pi}\,\bar D_{1} \,\Big[R_{i\bar j}\bar D_{2} \Phi^{\dagger\,\bar j}D_{1}\Phi^{i}\Big]=\frac{T_{G}g^{2}}{8\pi}\,\bar D_{1} \,{\cal T}_{21}
\,.
\end{split}
\end{equation}
In terms of classification of Ref. \cite{Dumitrescu:2011iu} it is the $R_{\,V}$ multiplet, $\partial_{11}{\cal T}_{22}
+\partial_{22}{\cal T}_{11}=0$.

\subsection{Hypercurrent in the heterotic {\boldmath ${\mathcal N}\!=(0,2)$} theory }
\label{hit}

As shown in Eq.\,(\ref{decomp}) transition to diminished ${\mathcal N}=(0,2)$ supersymmetry decomposes
the $ {\mathcal N}\!=(2,2)$ superfield $\Phi$ as
\beq
\Phi=A+\sqrt{2}\,\theta^{1}B\,,
\eeq
where the ${\mathcal N}=(0,2)$ superfields, introduced in Eq.\,(\ref{15}) and (\ref{15-1}), depend on $\theta^{\,2}=\theta_{R}$.
Correspondingly,
 the hypercurrent ${\cal T}_{\mu}$ decomposes
into two ${\mathcal N}\!=\!(0,2)$ supermultiplets,
\begin{eqnarray}
{\cal J}_{L}\!\!&=&\!\frac{1}{2}\,{\cal T}_{22}\Big |_{\theta^{1}=0}=\frac{1}{2}\,G_{i\bar j}\xbar{\!D}A^{\dagger\bar j}DA^{i}\,,
\nonumber
\\[3mm]
\widetilde {\mathcal T}_{RR}\!\!&=&\!\!-\frac{1}{2}\,\big[\xbar{\! D}_{1},D_{1}\big]{\cal T}_{11}\Big |_{\theta^{1}=0}=
{2}\,G_{i\bar j}\Big[\partial_{R}A^{\dagger\,\bar j}\partial_{R}A^{i}+i\,B^{\dagger \,\bar j}\nabla_{R}B^{i}
\Big] +{\rm H.c.}\,.
\label{jT}
\end{eqnarray}
These supermultiplets introduced in Ref.\cite{Dumitrescu:2011iu} have the following general structure,\footnote{
Our notations differ: ${\cal J}_{L}$ and $\widetilde {\mathcal T}_{RR}$ are  the same as $S_{++}$ and ${\cal T}_{----}$
in \cite{Dumitrescu:2011iu}.}
\begin{eqnarray}
{\cal J}_{L}\!\!&=& v_{L}+i\theta \,s_{L;L}
+i\theta^\dagger s_{L;L}^\dagger-\theta\theta^\dagger \vartheta_{LL}\nonumber
\,,\\[3mm]
\widetilde {\mathcal T}_{RR}\!\!&=&
\vartheta_{RR} +\,\theta \,\partial_{R}s_{R;L} -\theta^{\dagger}\partial_{R}s_{R;L}^{\dagger}
+\theta \theta^{\dagger}\partial_{R}^{2}v_{L}\,.
\label{jT1}
\end{eqnarray}

It is clear that the heterotic deformation does not change the expression for ${\cal J}_{L}$ but
modifies $\widetilde {\mathcal T}_{RR}$ supermultiplet where the lowest superfield component represents the $\vartheta_{RR}$ component of energy-momentum tensor. Namely, the expression for $\widetilde {\cal T}_{RR}$ in Eq.\,(\ref{jT}) is modified to
\beq
\widetilde {\mathcal T}_{RR}=\frac{1}{2}\left\{
G_{i\bar j}\Big[\partial_{R}A^{\dagger\bar j}\partial_{R}A^{i}\!+i\,ZB^{\dagger \bar j}\nabla_{R}B^{i}\!
 +\!i\kappa\,{\cal B}B^{i}\,\partial_{R}A^{\dagger \bar j}\Big] \!+i\,{\cal Z}{\cal B}^{\dagger}\partial_{R}{\cal B}\right\}+{\rm H.c.}
\,.
\eeq

Quantum anomalies for ${\cal J}_{L}$ and $\widetilde {\mathcal T}_{RR}$ according to \cite{Dumitrescu:2011iu}
have the following general form:
\beq
\begin{split}
\partial_{R}{\cal J}_{L}=\frac12 \,D\,\mathcal{W}-\frac12 \xbar {\!D} \xbar{\mathcal{W}}\,,\\[2mm]
\xbar{\!D}\,  \widetilde {\mathcal T}_{RR}=  \partial_{R}\mathcal{W}\,,\qquad\quad ~~
\end{split}
\label{Janom}
\eeq
where $\mathcal{W}$ represents the supermultiplets of anomalies,
\beq
\mathcal{W} = s_{R;L}^\dagger -i\theta\,\big(\vartheta_{LR}+{i}\,\partial_{R}\,v_{L}\big)-i\theta\theta^\dagger \partial_{L}s_{R;L}^\dagger\,.
\eeq

\begin{figure}[h]
\begin{center}
\includegraphics[width=5.7in]{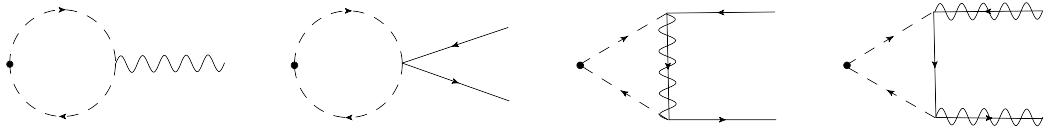}
\end{center}
\caption{\small One-loop diagrams for $v_{L}$ current. The dots denote the $v_{L}$ currents, dashed lines are quantum $A$ fields while wavy lines refer to the background $A$.}
\label{j2ano_111031a}
\end{figure}

At the one-loop level in the heterotically modified models $\mathcal{W}$ is the \nzt chiral superfield of the following form,
\beq
\mathcal{W}^{(1)}\!=\!\frac{1}{4\pi}\!\left\{R_{i\bar j}\,\partial_{R}A^{i}\xbar{\!D}A^{\dagger \bar j}-i\xbar{\!{D}}\left[Z(R_{i\bar j} -h^{2}G_{i\bar j})B^{\dagger \bar j}B^{i} -
d\,h^{2}{\cal Z}{\cal B}^{\dagger}{\cal B}\right]\right\}.
\label{Wanom}
\eeq
This expression can be verified through a component calculation of the one-loop graphs in Fig.\,\ref{j2ano_111031a}
for $v_{L}$ which is the lowest component of ${\cal J}_{L}$.
Rewriting Eq.\,(\ref{Wanom}) in terms of the heterotic curvature ${\cal H}$ defined in Eqs.\,(\ref{curv}, \ref{Hcomp}) we
arrive at
\beq
\mathcal{W}^{(1)}\!=\!\frac{1}{4\pi}\!\left\{R_{i\bar j}\,\partial_{R}A^{i}\xbar{\!D}A^{\dagger \bar j}-i\xbar{\!{D}}\left[R^{(B)}_{i\bar j}B^{\dagger \bar j}B^{i}-4{\cal H}_{ik\bar l}\xbar{\cal H}_{\bar j\bar k l}G^{(B)k\bar k} G^{l\bar l}B^{\dagger \bar j}B^{i}\right]\right\},
\eeq
where all right-moving fermions are included in $B^{i}$ (see Sec.\,\ref{geometry} for details).

Following Eqs.\,(\ref{LF}) and  (\ref{1loopL}) it is simple to verify that the superfield ${\mathcal W}$, which represents the supermultiplets of anomalies, coincides with the one-loop running of the superfield
\cf ~associated with the Lagrangian by Eq.\,(\ref{LF}),
\beq
{\mathcal W}^{(1)}=i\,M_{\rm uv}\frac{d}{d M_{\rm uv}}\,\cfe\,\big|^{\rm one-loop}\,.
\eeq

What about higher-loop corrections? They will show up as higher loops in the anomalous dimensions.
It means that Eq.\,(\ref{Wanom}) is modified to
\beq
\mathcal{W}\!=\!\frac{1}{4\pi}\Big[R_{i\bar j}\partial_{R}A^{i}\xbar{\!D}A^{\dagger \bar j}\!-i\xbar{\!{D}}\,\big(Z R_{i\bar j}B^{\dagger \bar j}B^{i}\big)\Big] +\frac{i}{2}\,\xbar{\!{D}}\,\Big[\gamma_{\psi_{R}}ZG_{i\bar j}B^{\dagger \bar j}B^{i}\! +\gamma_{\zeta}{\cal Z}{\cal B}^{\dagger}{\cal B}\Big].
\label{WanomH}
\eeq

\subsection{Analog of the Konishi anomaly and beta function}

Here we will discuss a relation between the hypercurrent anomalies and beta functions. It is an example of another
analog to 4D gauge theories. We mentioned above that the supermultiplet of anomalies ${\cal W}$ is given
by differentiation of the effective Lagrangian with respect to $\log M_{\rm uv}$. Let us  have a closer look at how it works for $\beta_{g}$\,.

At the one-loop level the running of the metric $dG_{i\bar j}/dL=R_{i\bar j}/2\pi$ is given by the Ricci tensor, see Eq.\,(\ref{1loopL}).
In ${\cal W}$ (see Eq.\,(\ref{WanomH})) it is represented by the term with the $A$ superfields. The terms with the $B$ fields contribute to the higher loops. We can simplify these terms using equations of motion plus possible anomalies. Using the equation of motion we get
\beq
\begin{split}
\xbar{\!D}\big( ZG_{i\bar j}B^{\dagger \bar j}B^{i}\big)\big |_{\rm class}=-\kappa \,G_{i\bar j} \xbar{\!D}A^{\dagger \bar j} B^{i}{\cal B}\,,
\\[2mm]
\xbar{\!D}\big( {\cal Z}{\cal B}^{\dagger}{\cal B}\big)\big |_{\rm class}=-\kappa\, G_{i\bar j} \xbar{\!D}A^{\dagger \bar j} B^{i}{\cal B}\,.
\end{split}
\label{eqm}
\eeq
There is also an anomalous part due to  loops in the background of the $A$ field, namely,
\beq
\xbar{\!D}\big( ZG_{i\bar j}B^{\dagger \bar j}B^{i}\big)\big |_{\rm anom}=-\frac{i}{4\pi} \,R_{i\bar j}\partial_{R}A
\xbar{\!D}A^{\dagger \bar j}\,.
\label{KonAn}
\eeq
This part  is a clear-cut analog of the Konishi anomaly in 4D \cite{KK}.

Using both, classical equations (\ref{eqm}) and the anomalous one (\ref{KonAn}), as well as $R_{i\bar j}= (T_{G}g^{2}/2) G_{i\bar j}$\,, we come from ${\cal W}$ of Eq.\,(\ref{WanomH}) to
\beq
{\cal W}=\frac{T_{G}g^{2}}{8\pi}\,G_{i\bar j}\partial_{R}A^{i}\xbar{\!D}A^{\dagger \bar j}\!\left(\!1\!-\! \frac{T_{G}g^{2}}{8\pi}\!+\!\frac{\gamma_{\psi_{R}}}{2}\right)
\!+\!\left( \frac{T_{G}g^{2}}{8\pi}\!-\!\frac{\gamma}{2}\right)\!i\kappa\,G_{i\bar j} \xbar{\!D}A^{\dagger \bar j} B^{i}{\cal B}\,.
\eeq
In the background of the $A$ field the integrating out right-moving fermions in the operator $G_{i\bar j} \xbar{\!D}A^{\dagger \bar j} B^{i}{\cal B}$
involves the same polarization operator $\Pi_{RR}$ as in Sec.\,\ref{sec:holo}, see (\ref{223}) and Appendix B, and results in
\beq
\Big\langle i\kappa\,G_{i\bar j} \xbar{\!D}A^{\dagger \bar j} B^{i}{\cal B}\Big\rangle_{\!\!A}=
\frac{h^{2}}{4\pi}\,G_{i\bar j}\partial_{R}A^{i}\xbar{\!D}A^{\dagger \bar j}\,.
\eeq
Thus, we get
\beq
\Big\langle {\cal W}\Big\rangle_{\!\!A}=\frac{1}{8\pi}\,G_{i\bar j}\partial_{R}A^{i}\xbar{\!D}A^{\dagger \bar j}\!\left[T_{G}g^{2}\left(\!1\!-\! \frac{T_{G}g^{2}}{8\pi}\!+\!\frac{\gamma_{\psi_{R}}}{2}\right)
\!+h^{2}\!\left( \frac{T_{G}g^{2}}{4\pi}\!-\!\gamma \right)\right]\,.
\label{avW}
\eeq

In the above calculations we limit ourselves by one loop  (besides higher loops in the anomalous dimensions $\gamma_{\psi_{R}}\,, \gamma_{\zeta}$\,). To see that the higher loops are needed it is sufficient to go to the (2,2) case when $h=0$\,. In this limit $\gamma_{\psi_{R}}=\gamma_{\zeta}=0$ but the the factor $1\!-\! (T_{G}g^{2}/8\pi)$ remains in (\ref{avW}). It should cancel out eventually for a pure bosonic field background.
Technically it happens in the following way. Accounting for the left-moving fermion anomaly in
$\partial_{R}\big(R_{i\bar j}\xbar{\!D}A^{\dagger \bar j}DA^{i})$ which cancels in the (2,2) case the one from right-movers leads to
a geometrical progression which turns the factor $1\!-\! (T_{G}g^{2}/8\pi)$ into $1/(1\!+\! (T_{G}g^{2}/8\pi))$. Then, in the bosonic
background this factor will be eaten up by integrating out left-moving fermions.

At nonvanishing $h$ one more geometrical progression is generated by  the factor \mbox{$1\!+\!(h^{2}/4\pi)$} which multiplies $T_{G}g^{2}$ in (\ref{avW}).
It is simple to understand this as just a summation of a chain of insertions of polarization operator $\Pi_{RR}$ into a bosonic propagator.
Thus, the multiloop expression for $\big\langle {\cal W}\big\rangle_{\!\!A}$ becomes
\beq
\Big\langle {\cal W}\Big\rangle_{\!\!A}^{\rm multi}=\frac{G_{i\bar j}\partial_{R}A^{i}\xbar{\!D}A^{\dagger \bar j}}{1+(T_{G}g^{2}/8\pi)}\,
\,\frac{T_{G}g^{2}\left(\!1\!+\!(\gamma_{\psi_{R}}/2)\right)-h^{2}\gamma}{8\pi(1-(h^{2}/4\pi))}
\,.
\label{avWm}
\eeq
The second factor in this expression is just $(-\beta_{g}/2g^{2})$ which gives the same $\beta_{g}$ as in Eq.\,(\ref{11}).
The normalzation follows from the one loop. The factor $1/(1+(T_{G}g^{2}/8\pi))$ will go away in the bosonic background
as it was explained above.

\section{Conclusion}
\label{conc}
\setcounter{equation}{0}

In this paper, we analyzed various quantum effects in the $\mathcal{N}\!=\!(0,2)$ deformed (2,2) two-dimensional sigma models.
The target spaces we considered generalize CP($N\!-\!1$) to the K\"ahler spaces which are
homogeneous spaces $G/H$.
 The $\mathcal{N}=(0,2)$ deformation
 thoroughly studied in this paper is called a nonminimal model.

We showed that  quantum effects will not deform geometry. Unlike other (0,2) models
there is no problem of internal anomalies when the heterotic modification is introduced 
the way we follow.  Another phenomenon closely related to  isometries  is integrability. One could ask whether the Lax relation holds, which would make these classes of models integrable, as in their undeformed cousins.

 The models we studied are characterized by two independent coupling constants.
 We analyzed them in perturbation theory. A crucial role belongs to the graph depicted in Fig.\,\ref{XXX}
 which is associated with the anomaly in the current that mixes the right-moving fermions. It is also anomalous
  in the sense that it produces a nonholomorphic contribution proportional to $|\kappa |^2$
 to the renormalization of $1/g^{2}$ at one loop. This effect then penetrates into higher orders.

 Using nonrenormalization theorems \cite{Cui:2011rz}, analogous to those in \cite{Novikov,shiva} in four-dimensional Yang-Mills, we derived a number of (perturbatively) exact relations between the $\beta$ functions and the anomalous dimensions of the fields $B$ and ${\mathcal B}$. Then we calculated the anomalous dimensions up to (and including) two loops thus obtaining explicit $\beta$ functions up to three loops.

 Then we studied the relation between the perturbative $\beta$ functions and the general hypercurrent analysis of Dumitrescu and Seiberg \cite{Dumitrescu:2011iu}. We found how the general structure of \cite{Dumitrescu:2011iu} is implemented in the nonminimal models under consideration. We demonstrated that the hypercurrent analysis leads to an alternative way for the $\beta$-function calculation provided that the two-dimensional analogs of the Konishi anomaly are
 taken into account.

Recently the $\mathcal{N}=(0,2)$ models attracted attention in connection with developments in the studies of surface operators in four dimensions (see, e.g., \cite{Gaiotto:2013sma}). In the nonminimal models  the protected quantities, i.~\!e., the chiral ring, are preserved under the (0,2) deformation. It is  interesting to pursue the calculation beyond the chiral sector exploring the nonchiral sector of the world sheet theory as a part of a 4D--2D coupled system. We hope that the results presented here can enlighten the very first step in pursuing such a goal.

Two-dimensional asymptotically free sigma models are long known to be excellent laboratories for modeling four-dimensional Yang-Mills theories.\ It was 40 years ago that A.\,Polyakov emphasized (in Ref. \cite{poly}) that asymptotically free two-dimensional sigma models
 could be the best laboratory for the four-dimensional Yang-Mills theories. His anticipation seems to be materializing.
 The nonminimal (0,2) sigma model discussed in this paper presents a close parallel to ${\mathcal N}\!=\!1$ super-Yang-Mills theory with matter in four dimensions (see also \cite{Cui:2011uw}).

\section*{Acknowledgments}
\addcontentsline{toc}{section}{Acknowledgments}

We thank T. Dumitrescu, A. Gadde, S. Gukov, K. Intriligator, A. Kapustin, Z. Komargodski, P. Koroteev, A. Losev, N. Nekrasov, G. Moore, H. Ooguri, 
E. Sharpe, A. Voronov, and A.\,Yung for helpful discussions.
X.C. would like to thank the European Organization for Nuclear Research, and the Galileo Galilei Institute (GGI) for
 hospitality. Parts of this project were worked out during X.C.'s stay  at the GGI Theoretical Physics workshop ``Geometry of Strings and Fields." The work of M.S. was supported in part by DOE Grant DE-FG02-94ER40823.  A.V. appreciates the hospitality of the Kavli Institute for Theoretical Physics where his research was supported in part by the National Science Foundation under Grant No.\ NSF PHY11-25915. 

\section*{Appendix A: Notation}
\label{notation}

\renewcommand{\theequation}{A.\arabic{equation}}
\setcounter{equation}{0}
\addcontentsline{toc}{section}{Appendix A}

We define the left-moving and right-moving derivatives as
\beq
\partial_{L}\equiv \partial_{LL}\equiv \partial_t+\partial_z\,,\qquad \partial_{R}\equiv
\partial_{RR}\equiv\partial_t-\partial_z\,.
\eeq
Correspondingly, the light-cone coordinates are
\beq
x_L = t-z\equiv x^0 - x^1\,,\qquad x_R = t+z \equiv x^0 + x^1\,.
\eeq
We use the following definition for the superderivatives:
\beq
D_L = \frac \partial {\partial \theta_R}-i\theta_R^\dagger \partial_{LL}\,,\qquad
\xbar D_L = -\frac\partial{\partial \theta_R^\dagger}+i\theta_R\partial_{LL}\,.
\label{a2}
\eeq
Their anticommutator gives $\{D_L,\xbar D_L\} = 2i\partial_{LL}\,$.

In the bulk of the paper we do not use $\theta_L$ and $D_R$. Hence we can omit the indices in (\ref{a2}),
\beq
\theta_R \to \theta\,,\qquad D_L \to D = \frac \partial {\partial \theta}-i\theta^\dagger \, \partial_{L}\,,\qquad
\xbar D_L \to \xbar{D}= -\frac\partial{\partial \theta^\dagger}+i\theta\, \partial_{L}\,.
\label{a3}
\eeq
We will consistently use the notation (\ref{a3}). Our normalization of the Berezin integral is
\beq
\int d\theta\,\theta =1\,,
\eeq
and
\beq
\int d^2\theta \equiv \int d\theta d\theta^\dagger\,.
\eeq

In passing from the ordinary to the light-cone coordinates  we must also change the components of Lorenz vectors,
tensors, etc. For instance,  for the supercurrent we have
\beq
s_{L;L}=s_{LLL}=(s^{0}_{L}+s^{1}_{L})/2\,,\qquad s_{R;L}=s_{RRL}=(s^{0}_{L}-s^{1}_{L})/2\,.
\eeq
Moreover, for the energy-momentum tensor  $T^{\mu\nu}$,
\beqn
T_{LL}=T_{LLLL}&=T_{00}+T_{10}+T_{11}+T_{01}\,,\nonumber\\[2mm]
T_{LR}=T_{LLRR}&=T_{00}+T_{10}-T_{11}-T_{01}\,,\nonumber\\[2mm]
T_{RL}=T_{RRLL}&=T_{00}-T_{10}-T_{11}+T_{01}\,,\nonumber\\[2mm]
T_{RR}=T_{RRRR}&=T_{00}-T_{10}+T_{11}-T_{01}\,.
\eeqn

\section*{Appendix B: Calculation of \boldmath{$\Delta_\kappa{\mathcal L}$}}
\label{app:polarization}

\renewcommand{\theequation}{B.\arabic{equation}}
\setcounter{equation}{0}
\addcontentsline{toc}{section}{Appendix B}

In this Appendix a detailed calculation of the crucial diagram presented in Fig.\,\ref{XXX}
is given.
In the coordinate space it proceeds as follows (the target space indices which go through are suppressed).
We start from the $|\kappa|^{2}$ correction to the action,
\beq
\Delta_\kappa S= \!\int\! d^{2}x \,\Delta_{\kappa}{\cal L}=\!\int\! d^{2}x\,d^{2}y \,\partial_{L}\phi^{\dagger \bar k}(x)G_{i\bar k}(x) \Pi_{RR}^{i \bar j}(x,y)G_{l\bar j}\partial_{L}\phi^{i}(y)\,,
\label{act_k}
\eeq
where the polarization operator and its expression via a Green function is defined in Eq.\,(\ref{223}). We choose
the background field $\phi^{i}$ in the form of the plane wave,
\beq
\phi^{i}(x)=f^{i}\,{\rm e}^{-ikx}\,,
\eeq
where $f^{i}$ are  constants. In such field the fermionic part of the action takes a form,
\beq
\begin{split}
S_{F}=\!\int\! d^{2}x &\Big[{\cal Z}\zeta_{R}^{\dagger}\left(1+ \frac{\partial_{\mu}\partial^{\mu}}{M^{2}}\right)i\partial_{L}\zeta_{R}
+ Z\psi^{\dagger \bar j}G_{i\bar j}\Big(i \delta^{i}_{k}\partial_{L} +\Gamma^{i}_{k} k_{L}\Big) \psi^{k}\\[2mm]
&+\big(\kappa {\rm e}^{-ikx} \zeta_{R}\psi_{R}^{i}G_{i\bar j}f^{\dagger \bar j}k_{L}+{\rm H.c.}\big) \Big ].
\end{split}
\eeq
Here $G_{i\bar j}=G_{i\bar j}(f,f^{\dagger})$ and $\Gamma^{i}_{k}=\Gamma^{i}_{lk}(f,f^{\dagger}) f^{k}$ are $x$-independent matrices.
Here we also introduced an UV regularization by higher derivatives in the propagator of $\zeta_{R}$. The Fourier transform of this  propagator
is
\beq
S_{\zeta}(p)=\frac{i}{{\cal Z}\,p_{L}} \,\frac{M^{2}}{M^{2}-p^{2}}\,.
\eeq
The Fourier transform of the $\psi_{R}$ propagator is
\beq
S_{\psi_{R}}^{i\bar j}=\frac{i}{Z} \left[\frac{1}{p_{L}\, I +k_{L}\,\Gamma}\right]^{i}_{k}G^{k\bar j}
\eeq
Then for the Fourier transform of the polarization operator $\Pi_{RR}^{i\bar j}$ we have
\beq
\Pi_{RR}^{i\bar j}(k)=ih^{2}\int \frac{d^{2}p}{(2\pi)^{2}} \,\frac{M^{2}}{M^{2}-p^{2}} \,\frac{1}{p_{L}}\,
\left[\frac{1}{(p_{L}+k_{L}\, I -k_{L}\,\Gamma}\right]^{i}_{k}G^{k\bar j}\,.
\eeq
It is simple to do an integration which results in
\beq
\Pi_{RR}^{i\bar j}(k)=-\frac{|\kappa|^{2}}{4\pi Z{\cal Z}}\left[\frac{K_{R}^{2}}{K_{\mu}K^{\mu}}\,\,\frac{\log(1-K_{\mu}K^{\mu}/M^{2})}{(-K_{\mu}K^{\mu}/M^{2})}\right]^{i}_{k}G^{k\bar j}\,,
\eeq
where we introduced the matrix
\beq
\left[K_{\mu}\right]^{i}_{k}=k_{\mu}\left[I-\Gamma\right]^{i}_{k}\,,
\eeq
representing the covariant derivative $i\nabla_{\mu}$\,.

For momenta $k\ll M$ the expression is simple,
\beq
\Pi_{RR}^{i\bar j}(k)=-\frac{|\kappa|^{2}}{4\pi Z{\cal Z}}\,\frac{k_{R}^{2}}{k_{\mu}k^{\mu}}\,G^{i\bar j}\,.
\eeq
Substituting this into Eq.\,(\ref{act_k}) we come to the result (\ref{corL}) for $\Delta_{\kappa} {\mathcal L}$\,.

The expression for $\Pi_{RR}^{i\bar j}(k)$ is related to anomaly in the polarization operator.
The  way we derived it could be called infrared, the $p$ integration was
contributed dominantly by $p\sim k$.  The ultraviolet derivation follows from
\beq
\left[K_{L}\right]^{k}_{i}\Pi_{RR}^{i\bar j}(k)=ih^{2}\int \frac{d^{2}p}{(2\pi)^{2}} \,\frac{M^{2}}{M^{2}-p^{2}} \left[\frac{1}{p_{L}}-
\frac{1}{p_{L}+K_{L}}\right]^{k}_{l}G^{l\bar j}\,.
\eeq
Integration here is dominated by $p\sim M$ and gives for $k\ll M$,
\beq
\left[K_{L}\right]^{k}_{i}\Pi_{RR}^{i\bar j}(k)=-\frac{|\kappa|^{2}}{4\pi Z{\cal Z}} \left[K_{R}\right]^{k}_{l} G^{l\bar j}\,,
\eeq
which corresponds to Eq.\,(\ref{UVk}) in the text.

\end{document}